\documentclass{article}
\usepackage{arxiv}

\usepackage{graphicx,amsmath,amsfonts,amscd,amssymb,pstricks}
\usepackage{graphicx,color}
\usepackage{dcolumn}
\usepackage{bm}
\usepackage{longtable}
\usepackage{amsmath}
\usepackage{mathrsfs}
\usepackage{amsfonts}
\usepackage{textcomp}
\usepackage{esvect}
\usepackage{float}
\usepackage[USenglish,british]{babel}
\usepackage{environ}
\usepackage{xifthen}
\usepackage{xargs}			
\usepackage{hyperref}
\usepackage{physics}
\usepackage{multirow,here}

\title{Analysis of the non-linear beam dynamics at top energy for the CERN Large Hadron Collider by means of a  diffusion model}
\author{A.~Bazzani \\
Physics and Astronomy Department, Bologna University and INFN-Bologna \\
\And
M. Giovannozzi\thanks{Corresponding author: massimo.giovannozzi@cern.ch} \\
Beams Department, CERN, 1211 Geneva 23, Switzerland \\
\And 
E.H~Maclean \\
Beams Department, CERN, 1211 Geneva 23, Switzerland \\
University of Malta, Msida, Malta}
\begin{document}
\maketitle
\begin{abstract}
In this paper the experimental results of the recent dynamic aperture at top energy for the CERN Large Hadron Collider are analysed by means of a diffusion model whose novelty consists of deriving the functional form of the diffusion coefficient from Nekhoroshev theorem. This theorem provides an optimal estimate of the remainder of perturbative series for Hamiltonian systems. As a consequence, a three-parameter diffusion model is built that reproduces the experimental results with a high level of accuracy. A detailed discussion of the physical interpretation of the proposed model is also presented.
\end{abstract}
\section{Introduction}
The dynamic aperture (DA) is the amplitude of the phase space region where stable, i.e. bounded, motion occurs. DA is one of the key quantities for the design of modern colliders based on superconducting magnets, which feature unavoidable non-linear field errors, such as the Tevatron~\cite{tev1,tev3,tev4}, HERA~\cite{herap1,herap2,herap3,herap4}, RHIC~\cite{rhic}, the Superconducting Super Collider (SSC)~\cite{ssc1,ssc2}, and LHC (see, e.g. Ref.~\cite{LHCDR} for a detailed overview). 

The study of transport in the phase space of non-integrable Hamiltonian systems is a very difficult problem due to the coexistence of weakly-chaotic regions and invariant Kolmogorov--Arnold--Moser (KAM) tori~\cite{KAM} that implies a sensitive dependence of the orbit evolution on the initial conditions. The relevance of Arnold diffusion~\cite{arnold}, a generic phenomenon in Hamiltonian systems with two or more degrees of freedom, in  applications is still debated. 

Macroscopic physical systems cannot realise the symplectic character of the dynamics at arbitrary spatial and time scales. Nonetheless, some results of Hamiltonian perturbation theory turn out to be robust with respect to the details of the considered system and they can provide effective laws for the study of stability and diffusion problems of the orbits. Nekhoroshev  theorem~\cite{nekhor} is an excellent example of such a result, the corresponding estimate for the orbit stability time being applied in several fields ranging from Celestial Mechanics to Accelerator Physics, { where  in recent years, a connection between Nekhoroshev theorem and time variation of the dynamics aperture has been established\cite{dacomp}.}

In a mathematical sense, the stability property requires an arbitrarily large time scale. In a physical context, however, particle stability can be linked to a maximum number of turns $N_{\rm max}$ that is determined on the basis of the specific application. Let $(x,y)$ be the transverse spatial coordinates describing the betatronic motion in a collider, if an ensemble of initial conditions defined on a polar grid ($x = r \, \cos \theta \, , \,\, y = r \, \sin \theta \,\,\, 0 \leq \theta \leq \pi/2$, where $x,y$ are expressed in units $\sigma_x, \sigma_y$ of the beam dimension) is tracked for up to $N_{\rm max}$ turns, then a measure of the DA can be defined as~\cite{dacomp}:
\begin{equation}
D(N) = \frac{2}{\pi} \int_0^{\pi/2} r(\theta;N) \, d\, \theta \equiv 
\langle  r(\theta; N) \rangle \, .
\label{DAdef}
\end{equation}
where $r(\theta; N)$ stands for the last stable amplitude in the direction $\theta$ for up to $N$ turns. Note that in case the stable phase space region is made of disconnected parts, only the area surrounding the origin is retained in these computations. In this way, the DA can be considered a function of $N$, with an asymptotic value, when it exists, representing the DA for an arbitrary large time. 

An accurate numerical computation of DA, as well as a good estimate of the numerical error associated with the numerical protocol used is of paramount importance to ensure the reliability of DA as a figure-of-merit for assessing synchrotron performance. A general discussion of the DA definition, its computation, and accuracy can be found, e.g. in Ref.~\cite{dacomp}. 

DA  computation requires the determination of the evolution of a large number of initial conditions, distributed to provide good coverage of the phase space under study, to probe whether their motion remains bounded over the selected time interval. { While the computational burden of a large set of initial conditions can be easily mitigated by means of parallelism~\cite{parallel}, it is not possible to mitigate the heavy CPU power needed for long-term simulations.} Hence, studies have explored the possibility to describe the DA dependence on the number of turns using simple models~\cite{dynap1,dynap2}. The underlying idea is that long-term behaviour of the DA can be extrapolated using knowledge from numerical simulations performed over a smaller number of turns. Additionally, a more efficient estimate of the long-term behaviour of the DA would expedite analysis of several configurations of the circular accelerator, which is sometimes mandatory to gain insight in the deeper nature of the beam dynamics. 

The Nekhoroshev~\cite{nekhor} theorem suggests an answer to the quest for modelling the time-evolution of DA. In fact, according to the results of Refs.~\cite{dynap1,dynap2}, the following scaling law holds
\begin{equation}  
D(N) = D_{\infty} + \frac{b}{ \left ( \log N \right )^{\kappa}} \, ,
\label{main}
\end{equation}
where $D_{\infty}$ represents the asymptotic value of the amplitude of the stability domain, $b$ and $\kappa$ being additional parameters. 

The model~(\ref{main}) gives the following rough description of the transverse phase space, in which we distinguish three macroscopic regions: an inner central core around the origin $r < D_{\infty}$, where the measure of the KAM~\cite{KAM} invariant tori is large, thus producing a stable behaviour apart from a set of a very small measure where Arnold diffusion can take place; a surrounding region, with $r > D_{\infty}$, where a weak chaos is present and the escape rate is reproduced by a Nekhoroshev-like estimate~\cite{nekhor,gtnekhor,gtnekhor1}; an outer region where most of orbits escapes quickly towards the infinity. In the region $r > D_{\infty}$ the model~(\ref{main}) provides an estimate of the stability time as a function of the amplitude $r$ of the form 
\begin{equation}
N(r) \, = \, N_0 \, \exp \left [{\left ( \frac{r_\ast}{r} \right )^{1/\kappa}} \right ]
\label{nekstab}
\end{equation}
where $N(r)$ is the number of turns that are estimated to be stable for particles with initial amplitude smaller than $r$. 

According to the scaling law~(\ref{main}), it has been proposed a model for the evolution of beam intensity in a hadron synchrotron~\cite{DAlosses}, which is the basis of the novel experimental method used to probe DA. 

In this paper, { we use a diffusive approach to reproduce the experimental results from the recent DA experiment at top energy in the LHC.} The beam dynamics in the weakly-chaotic region is governed by a stochastically-perturbed Hamiltonian system, which in turns is described by means of a Fokker-Planck (FP) equation~\cite{hamstoc}, whose solution represents the average evolution of the beam distribution, including also absorbing boundary conditions. This approach allows the diffusion process for the particle distribution to be simulated, providing a natural description of the beam dynamics in the presence of a collimation system, which is a typical situation in colliders based on superconducting magnets. 

{ The novelty of the proposed approach consists of using the remainder estimate of the perturbative series from Nekhoroshev theorem as functional form for the diffusion coefficient of the FP equation.} Indeed, Nekhoroskev approach to the optimal estimate of the remainder of perturbative  series~\cite{gtnekhor} represents the link between the analysis of the beam dynamics based on the scaling law of DA and that based on a diffusion equation: for the first the theorem provides the form of the scaling law of $D(N)$, while for the latter the theorem provides the form of the diffusion equation. 

The plan of the paper is the following: in section~\ref{sec:diffusion} the main aspects of the theory of diffusion processes in stochastically-perturbed Hamiltonian system are reviewed, while in section~\ref{sec:exp} the experimental technique is described. The main results of our analysis are presented and discussed in section~\ref{sec:anal}, where the detailed comparison between the theoretical approach based on the diffusion equation and the experimental measures is presented. In section~\ref{sec:tracking} the phase space of the system under consideration is studied by means of symplectic-tracking simulations { to provide a confirmation of the assumptions used for the diffusive approach}. Some conclusions are drawn in section~\ref{sec:conclusions}, { whereas the mathematical details of the proposed approach are presented in Appendices~\ref{sec:app1}, \ref{sec:app2}, \ref{sec:app3}.}
\section{Theoretical background}\label{sec:diffusion}
The results of perturbation theory of Hamiltonian systems imply that when the set of invariant KAM tori in phase space has a large measure, the orbits diffusion is possible only for a set of initial conditions of extremely small measure~\cite{morbid95}. Therefore, the existence of macroscopic diffusion phenomena in phase space has to be related to the presence of weak chaotic regions of large measure in which the large majority of KAM tori are broken~\cite{froes99}. { Note that in realistic models of betatron motion, slow modulation of the strength of lattice elements, transverse tune ripple induced by synchrotron motion, or weak stochastic effects, such as noise in active devices, may lead to the appearance of such regions.}

Nekhoroshev's theorem provides optimal estimates for the remainders of the asymptotic perturbative series { for Hamiltonian flows}, however, also in the case of a symplectic map in the neighbourhood of an elliptic fixed point, it is possible to provide an optimal estimate for the Birkhoff normal forms series~\cite{gtnekhor1,gtnekhor}.

Let $I$ be the unperturbed action, there exists an optimal perturbation order of the Birkhoff's expansion at which the remainder is estimated according to
\begin{equation}
\Vert R \Vert = A \exp\left [-\left (\frac{I_\ast}{I}\right )^{1/(2\kappa)} \right ]
\label{nek0}
\end{equation}
where $I_\ast$ represents an apparent radius of convergence of the perturbative series and the exponent $\kappa$ depends on the number of degrees of freedom of the system under consideration. We conjecture that under generic conditions the functional form of Nekhoroshev estimate~\eqref{nek0} can be applied to measure the strength of the chaotic component of the dynamics in the weak chaotic region. 
{ Assuming a diffusion approach for the evolution of the action distribution (see Appendices for the mathematical details) in the one-dimensional case the Fokker-Planck equation holds}
\begin{equation}
\frac{\partial \rho}{\partial t}=
\frac{\varepsilon^2}{2}\frac{\partial}{\partial I}\mathcal{D}(I)
\frac{\partial}{\partial I}\rho(I,t) \, .
\label{fokker2}
\end{equation}
where $\varepsilon$ is a scaling factor related to the perturbation amplitude. The Nekhoroshev's estimate suggests that the following functional form for the action-diffusion coefficient
\begin{equation}
\mathcal{D}(I)=c \, \exp\left [-2 \left (\frac{I_\ast}{I}\right )^{1/(2\kappa)}\right ]
\label{diffnek}
\end{equation}
is suitable to simulate the action diffusion under the previous assumptions. The constant $c$ is computed by normalising the diffusion coefficient according to 
\begin{equation}
c \, \int_0^{I_{\rm abs}}  \exp\left [-2 \left (\frac{I_\ast}{I}\right )^{1/(2\kappa)}\right ]dI=1 \, ,
\label{normal}
\end{equation}
where $I_{\rm abs}$ represents the position of the absorbing boundary condition. The physical meaning of the parameters $(\varepsilon, \kappa, I_\ast)$ that characterise the diffusion model~\eqref{fokker2} and~\eqref{diffnek} is readily derived from Nekhoroshev's theorem: i) $\varepsilon$ is an adimensional quantity that measures the strength of the non-linear effects acting on the beam; ii) the exponent $\kappa$ emerges from the analytic structure of the perturbative series and it mainly depends on the phase-space dimensionality and on the nature of the non-linear terms { that occur in the perturbative series, independently from their strength}; iii) $I_\ast$ { reveals the asymptotic character of the perturbative series and it is related to the strength of non-linear terms. Usually, the region in phase space corresponding to $I\simeq I_\ast$ is beyond the short-term dynamic aperture, where our approximation is no more valid.} It is worthwhile mentioning that the parameters $\varepsilon$ and $I_\ast$ are in principle correlated since a scaling in the action changes the strength of the perturbation. However, the position of the absorbing barrier is invariant with respect to the global time scaling $\varepsilon$, whereas it depends on the action scaling.

\begin{figure}[htb]
\centering
\begin{tabular}{@{}c@{}}
{\includegraphics[trim= 5mm 60mm 20mm 50mm, width=0.6\linewidth,clip=]{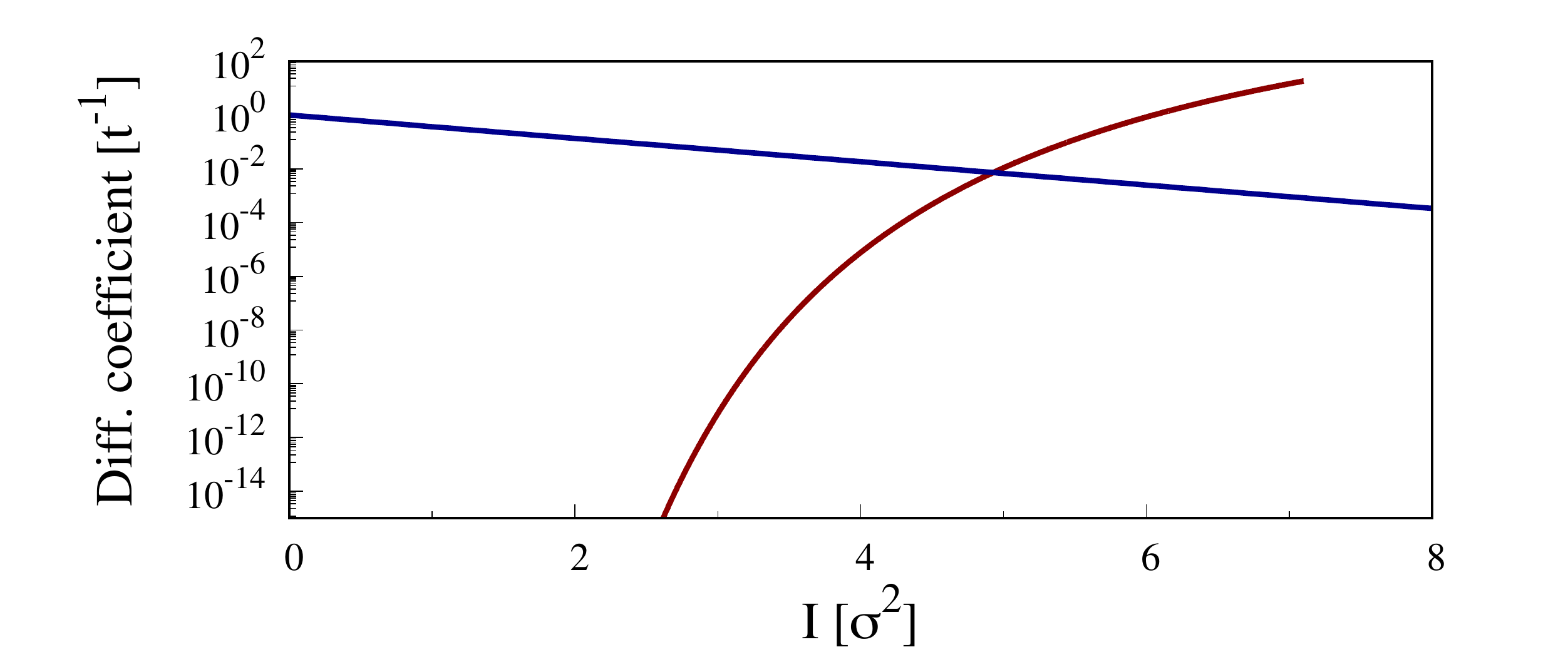}} \\
{\includegraphics[trim= 5mm 60mm 20mm 50mm, width=0.6\linewidth,clip=]{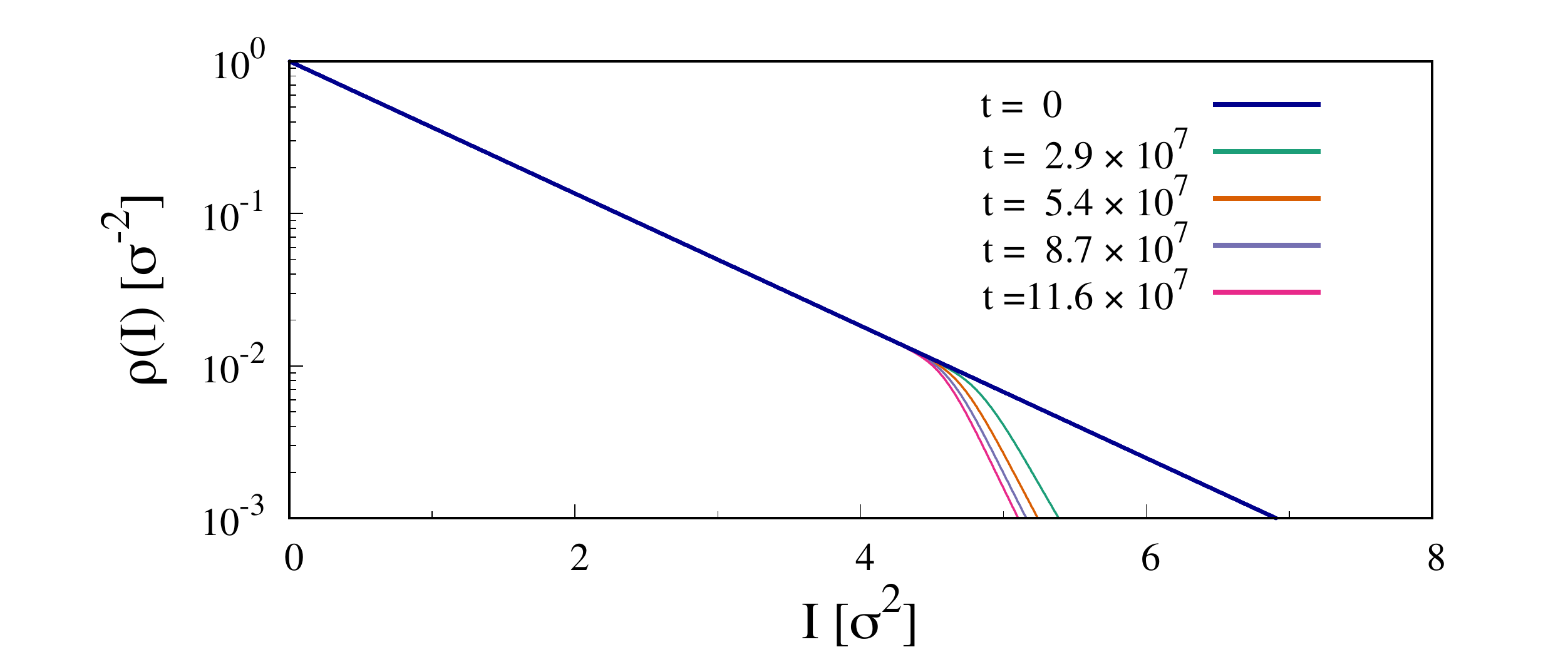}}
\end{tabular}
\caption{Upper: plot of the Nekhoroshev diffusion coefficient~\eqref{diffnek} (red curve) as a function of $I$ for $\kappa=0.33$ and $I_\ast=21.5$. To highlight the features of the function~\eqref{diffnek} the exponential $\exp(-I)$ (blue curve) is also shown. Lower: evolution of the 1D transverse beam distribution using the FP equation~\eqref{fokker2} with the Nekhoroshev diffusion coefficient of the upper plot. The parameter values used in the simulations correspond to the experimental set-up for Beam 2 with correctors V blow-up (see Table~\ref{summary_table}). }
\label{fpfig}
\end{figure}

In Fig.~\ref{fpfig} we plot the behaviour of the diffusion coefficient~\eqref{diffnek} (upper) using parameter values relevant for the comparison with experimental data (see section~\ref{sec:anal}) and we show an example of the numerical solution of the FP equation~\eqref{fokker2} (lower) with exponential initial distribution and an absorbing boundary condition. 

One clearly sees the effects of the functional form of the Nekhoroshev diffusion coefficient: after a rather fast initial diffusion, the evolution of the beam distribution slows down. Moreover, the changes to the initial shape of the distribution are limited to its tails. This is the reason for the existence of a stable region of finite extent in phase space for finite times, which would give rise to a finite dynamic aperture. 
\section{LHC dynamic-aperture experiment at top energy} \label{sec:exp}
{ With the advent of the LHC and the approval of its high-luminosity upgrade~\cite{TDR} the topic of measuring the DA by means of beam experiments, has regained interest, after a break between the design phase of the LHC (see, e.g. Refs.~\cite{meas1,meas2,meas3} and Ref.~\cite{comp} for a review of the comparison between measurements and simulations), and its commissioning and following operation periods.}

\begin{figure}[H]
\begin{center}
\begin{tabular}{@{}c@{}}
{\includegraphics[trim = 4mm 0mm 5mm 2mm, width=0.5\linewidth,clip=]{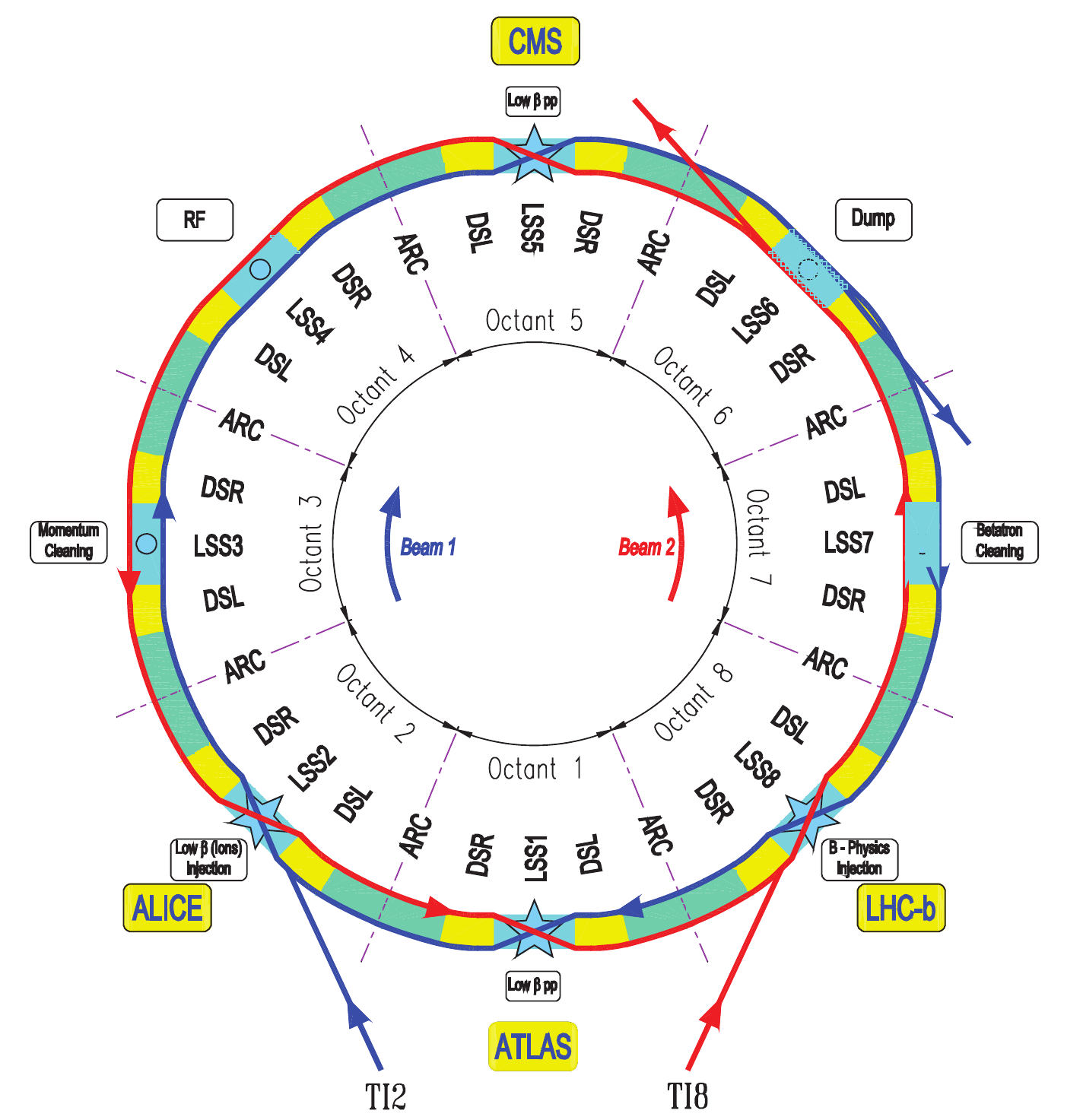}} \\
{\includegraphics[width=0.5\linewidth,clip=]{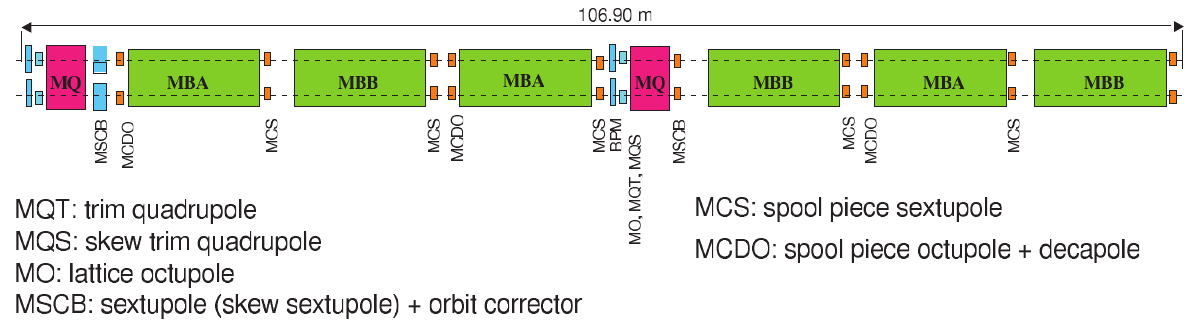}} \\
{\includegraphics[trim = 0mm 0mm 0mm 45mm, width=0.5\linewidth,clip=]{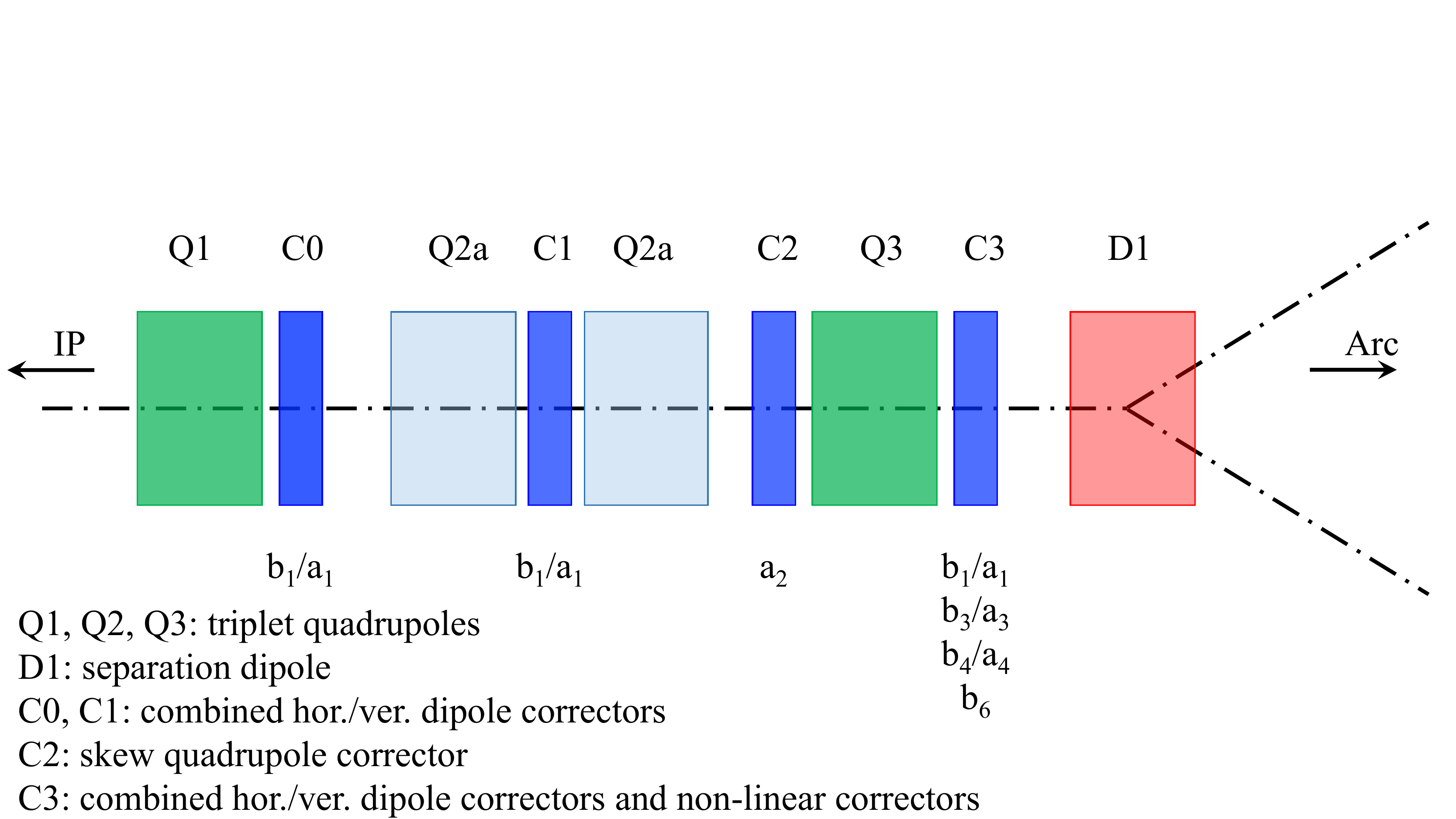}}
\end{tabular}
\caption{Upper: Layout of the LHC (from Ref.~\cite{LHCDR}). The ring eight-fold symmetry is visible, together with the arcs and the long straight sections. Middle: Layout of the LHC regular cell (from Ref.~\cite{LHCDR}). Six dipoles and two quadrupoles with the dipole, quadrupole, sextupole, and octupole magnets (for closed orbit, tune, chromaticity correction and beam stabilisation, respectively) are shown. The spool pieces used to compensate the systematic $b_3$ component (MCS), $b_4$ and $b_5$ components (MCDO nested magnets) are also shown. Bottom: Sketch of the layout of the inner triplets and the non-linear correctors used in the experimental tests reported in this paper. The field imperfections of LHC magnets are represented as $B_y + i \, B_x = B_{\rm ref} \sum_{n=1}^{M} \left ( b_n  + i \, a_n\right ) 
\left ( \frac{x + i \, y}{R_{\rm r}} \right )^{n-1}$ where $R_{\rm r}=17$~mm.}
\label{LHClayout}
\end{center}
\end{figure}

DA measurements at the LHC (see Fig.~\ref{LHClayout}, upper, for a layout of the LHC ring) have been already carried out at injection energy~\cite{DABeam2,DABeam1_1,DABeam1_2} using different approaches, i.e. the standard kick method~\cite{DABeam2} or the new approach~\cite{DABeam1_1,DABeam1_2}. For the latter, the technique consists of blowing up both the horizontal and vertical emittances until beam losses can be detected. The beam intensity as a function of time is then recorded, fitted, and compared with the results of numerical simulations, usually showing a very nice agreement~\cite{DA_PRAB}. During these measurements, the strength of non-linear elements located in the regular cell of the accelerator (see Fig.~\ref{LHClayout}, middle, for a layout of the LHC cell, including also the non-linear correctors used) is varied to provide several machine configurations to be studied by means of numerical simulations. The encouraging results obtained at injection energy suggested to pursue the DA measurement at flat top energy.

The goals of the DA measurements performed at $6.5$~TeV in the LHC were many-fold: the use of squeezed optics allows probing the impact on beam dynamics of the non-linear field errors stemming from the quadrupoles in the high-luminosity insertions. Thus, one could examine and quantify the influence on beam loss and lifetime from changes in the strength of the normal dodecapole correctors (see Fig.~\ref{LHClayout}, bottom, for a sketch of the high-luminosity insertions, whose magnets were used during the experiment) in the ATLAS and CMS interaction regions (IR) 1 and 5, respectively. This aspect is particularly relevant in view of the future High Luminosity LHC project~\cite{TDR}, for which the operational strategy to set the non-linear correctors in the high-luminosity IRs is still to be studied. Moreover, in previous studies, the beam emittance was heated to large values in both horizontal and vertical planes. While this can be considered a benefit of the method in-so-far as it gives a measure of changes to the average DA over all angles in the $x-y$ plane, it may also be regarded as a limitation since with such an approach it is not possible to distinguish between changes of the horizontal and vertical dynamic aperture. To help rectify this aspect, it was decided to measure the dynamic aperture simultaneously for three bunches in a single beam. One bunch was heated horizontally ({ H blow up, in the following}), one vertically ({ V blow up, in the following}), and one in both planes ({ H-V blow up, in the following}). Note that a witness bunch of small transverse emittance provides a reference case. The key objective of these measurements was related to the time scale achieved. Typical DA simulations are performed over $10^{5}-10^{6}$~turns ($\sim 8-88$~s of LHC operation) and previous measurements have been performed on the $5-10$~minute time scale~\cite{DABeam1_1,DABeam1_2}. Operational time scales at top energy in the LHC, by contrast are of the order of $\sim 12$~h. To justify the extrapolation of simulated data that can be viably studied numerically to orders of magnitude longer times, it is also necessary to establish whether the analytical scaling laws hold over these same time scales. Thus the final objective of this novel measurement campaign was to perform dedicated dynamic aperture measurements on the time scale of an hour, significantly longer than any previous measurement in the LHC.

The experiment was performed using both the clockwise beam (Beam~1) as well as its counter-clockwise partner (Beam~2). The first was made of a single bunch blown up in both horizontal and vertical planes, while the latter comprised four bunches with different emittance blow as mentioned earlier. The transverse damper was used to provide a dipolar excitation, which blows up the transverse emittance due to band-limited, white noise excitation that is injected into the transverse damper feedback loop~\cite{ADT}. 

The value of $\beta^{*}$ in the IR1 and 5 experimental insertion was $0.4$~m. The primary collimators were set at $\sim 9\,\mathrm{\sigma_{\rm nom}}$, while the tertiary collimators were positioned at $\geq 15\,\mathrm{\sigma_{\rm nom}}$, which are significantly in excess than that defined by the horizontal and vertical primaries. The value of $\sigma_{\rm nom}$ is computed assuming the nominal value of the rms normalised emittance, namely $\epsilon^{*}_{\rm nom}=3.75\,\mathrm{\mu m}$.

After removing large orbit bumps in the experimental insertions, the fractional tunes were re-corrected to $(0.31,0.32)$ and chromaticity was set to $Q'_{x,y}=3.0$ units for both planes and beams. Linear coupling was trimmed down to a value of $|C^{-}|\approx 0.001$, which is at the limit of the measurement resolution. Having established the baseline conditions for the study, DA measurements were first performed by aggressively blowing up the Beam~2 bunches using the transverse damper up to very large emittances $\sim~25~\mathrm{\mu m}$. Large dodecapole sources were introduced by powering the IR-$b_{6}$ correctors left and right of the interaction point (IP) 1 and 5 uniformly to their maximum current. Then, the single bunch in Beam~1 was also blown up in horizontal and vertical planes, thus allowing DA measurements for both beams. Approximately $1$~hour of intensity data was recorded in this configuration. Finally the IR non-linear corrections for normal and skew sextupole and normal and skew octupole errors, which had been commissioned at the start of 2017, were collectively removed, and approximately $30$~minutes of intensity data was recorded for this final configuration. Additional details regarding the experimental session and the LHC set-up can be found in Ref.~\cite{MD_note}.

\begin{figure}[htbp]
\centering
\begin{tabular}{@{}c@{}@{}c@{}@{}c@{}@{}c@{}}
{\includegraphics[trim = 20mm 20mm 70mm 50mm, width=0.25\linewidth,clip=]{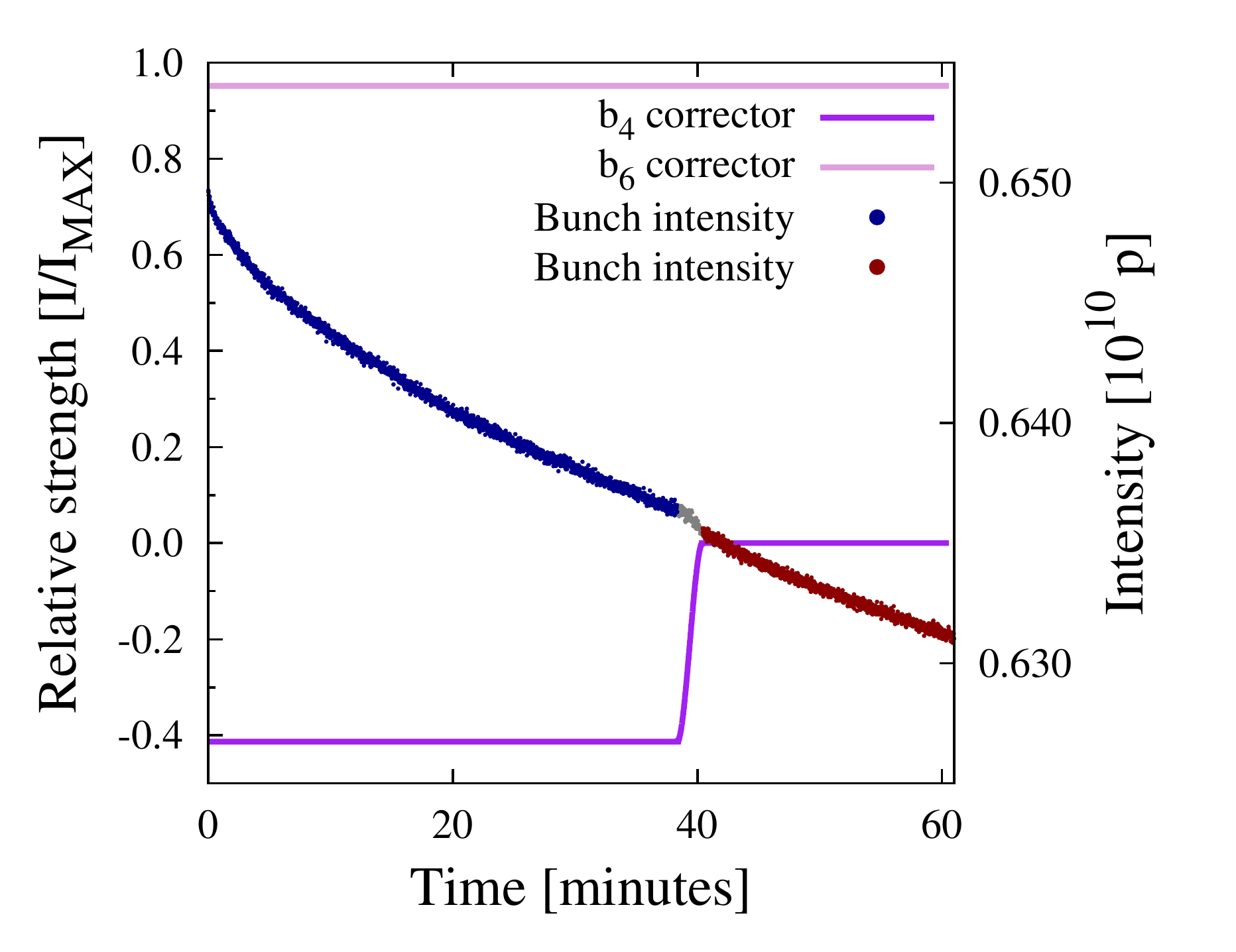}} &
{\includegraphics[trim = 20mm 20mm 70mm 50mm, width=0.25\linewidth,clip=]{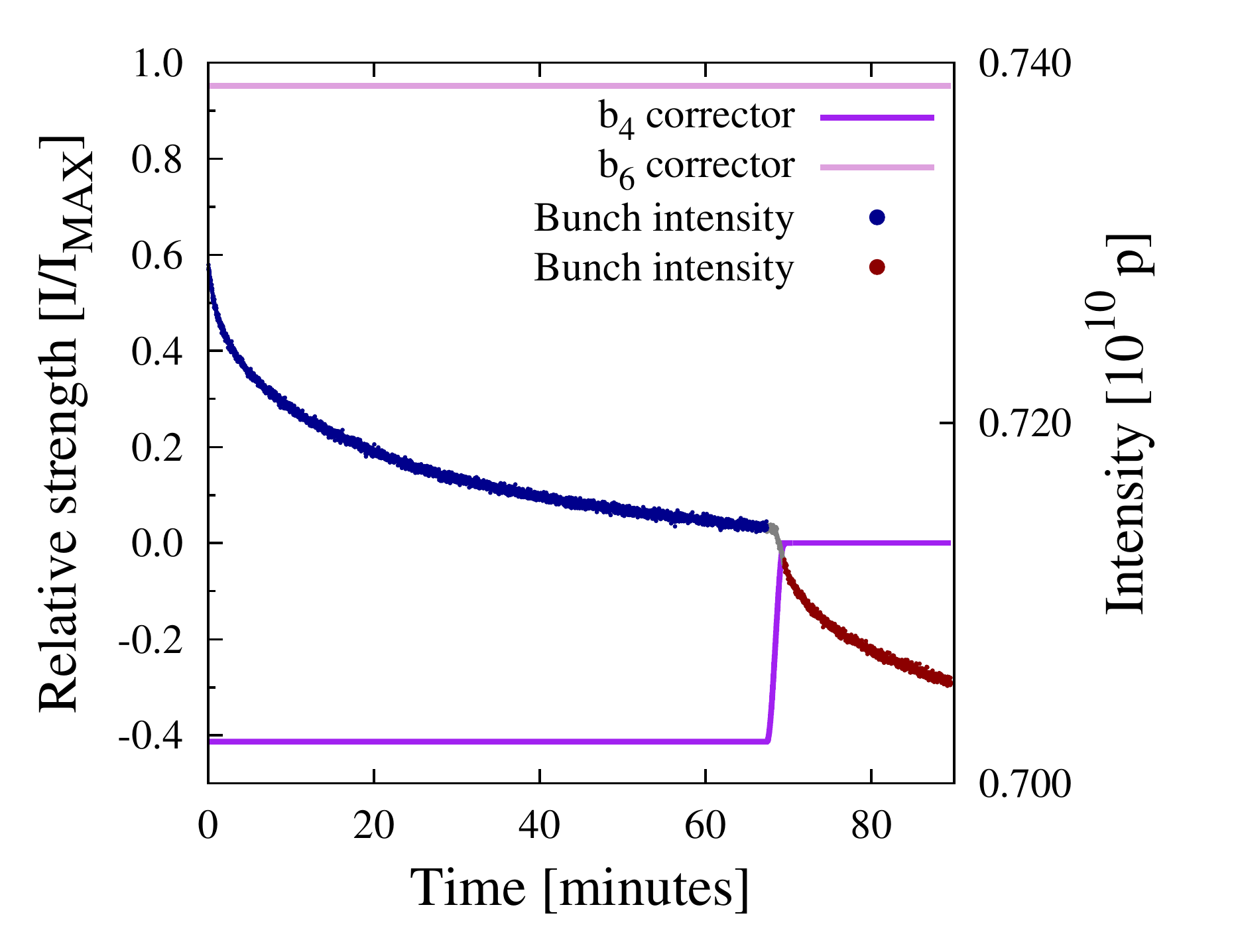}} &
{\includegraphics[trim = 20mm 20mm 70mm 50mm, width=0.25\linewidth,clip=]{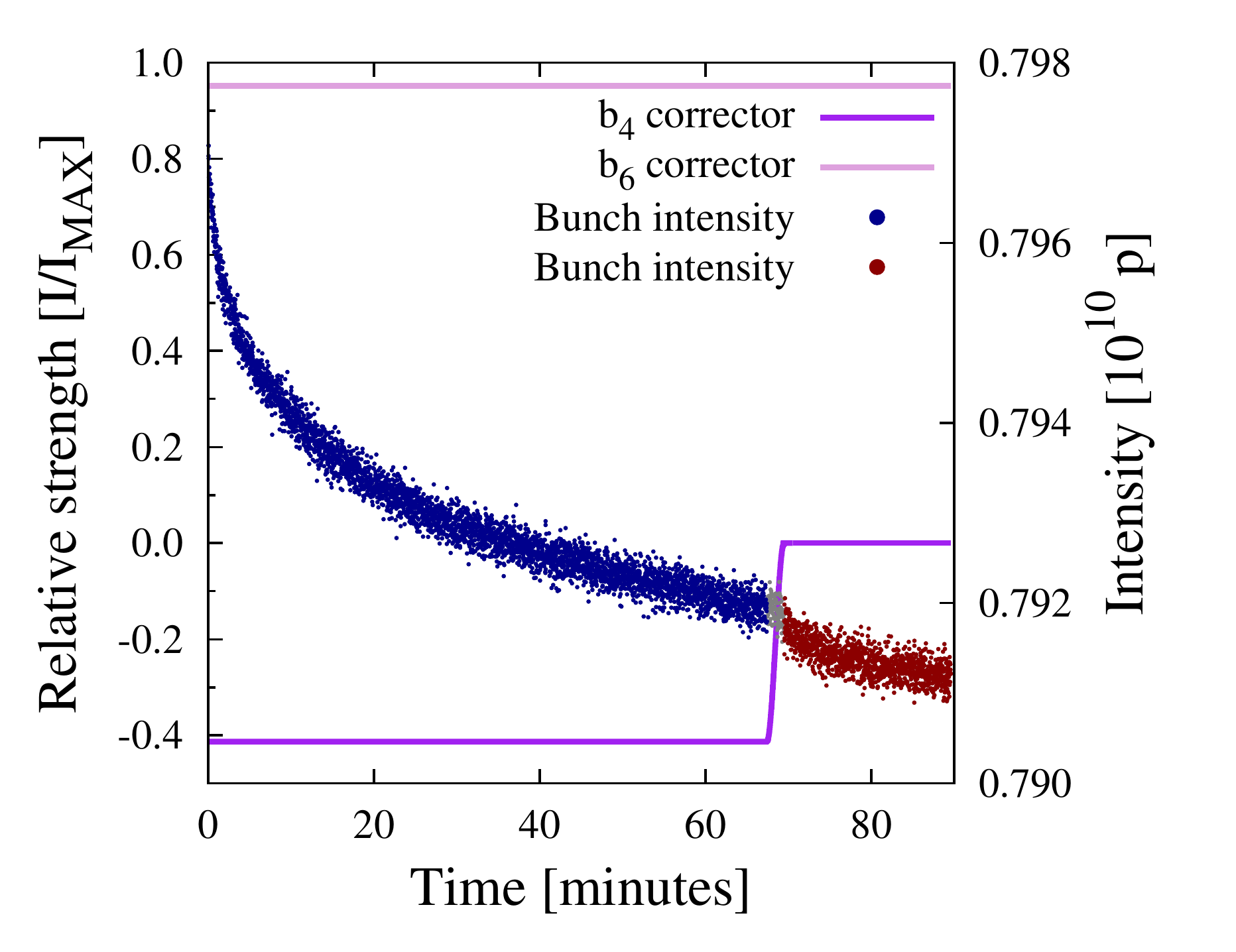}} &
{\includegraphics[trim = 20mm 20mm 70mm 50mm, width=0.25\linewidth,clip=]{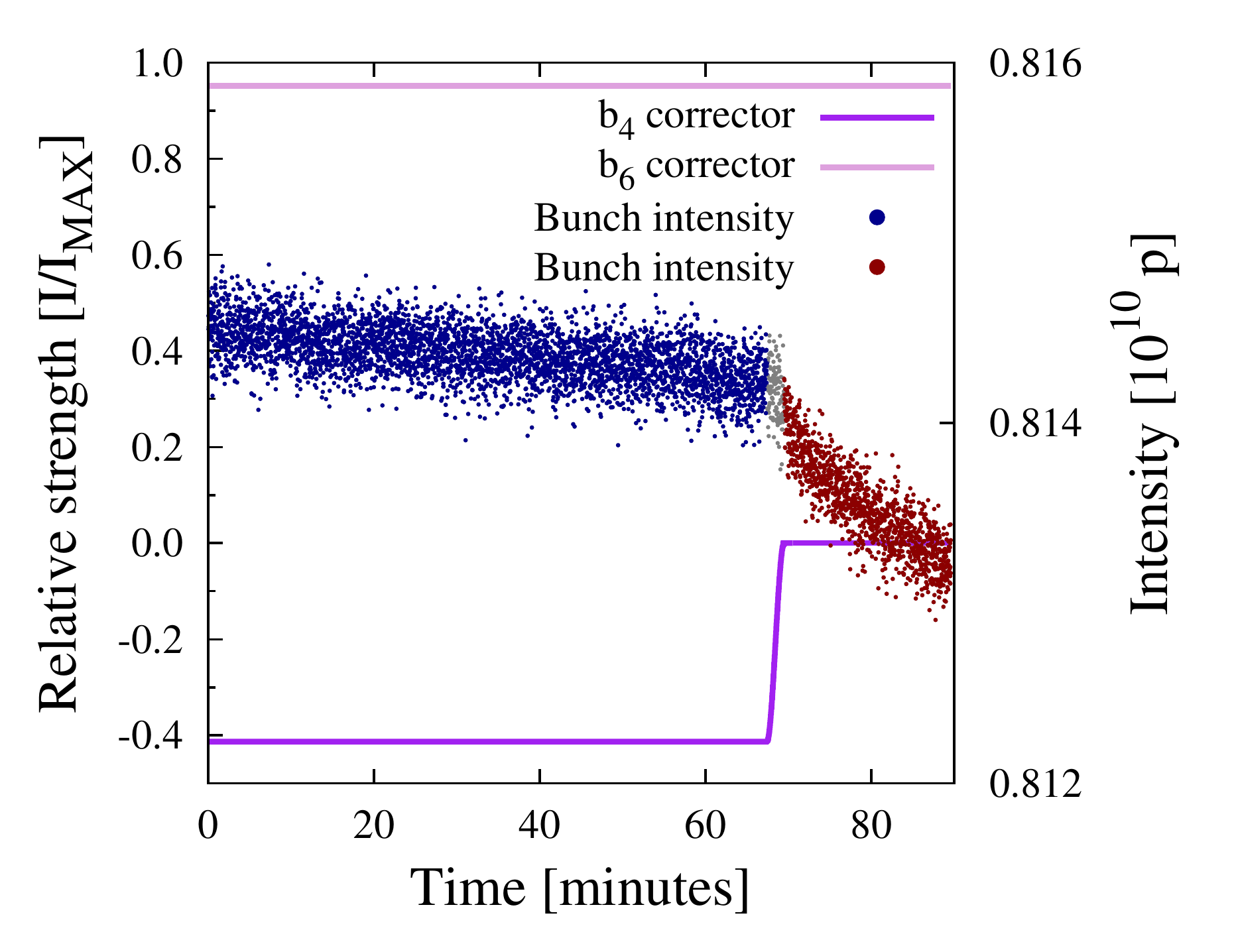}} \\
(a) & (b) & (c) & (d) \\
\end{tabular}
\caption{Summary plots of the DA measurements performed at $6.5$~TeV. The time-variation (note the different origin of the time scal for Beam~1 and 2) of the relative strength of the non-linear circuits used for the tests are shown, together with the bunch intensity evolution. The strength of the $b_4$ corrector is given as an example of the time variation for the $a_3, b_3, a_4$  correctors. The blue and red regions of the intensity curves are used in the numerical simulations, while the grey regions are discarded as they correspond to the transient state during the strength variation of the non-linear circuits. The results are: (a) H-V blow up, Beam~1; (b) H-V blow up, Beam~2; (c) V blow up, Beam~2; (d) H blow up, Beam~2.}
\label{summary_plot}
\end{figure}

The summary plots from the experimental session are shown in Fig.~\ref{summary_plot}, where the evolution of the relative strength of the non-linear correctors and the bunch intensity are visible. The two machine configurations are characterised by different levels of beam losses depending on the transverse emittances. A careful inspection of the summary plots for Beam~2 leads to the conclusion that the beam losses occur preferentially in the vertical plane. It is also worth stressing that the two beams are coupled by the single-aperture magnets in the experimental IRs, this is, e.g. the case of the non-linear correctors used in this experiment, whereas the remaining parts of the two rings are different, which implies that a different behaviour of Beam~1 and 2 for similar conditions of emittance blow up should not come as a surprise.

\begin{figure}[htb]
\centering
{\includegraphics[trim = 10mm 50mm 10mm 50mm, width=0.6\linewidth,clip=]{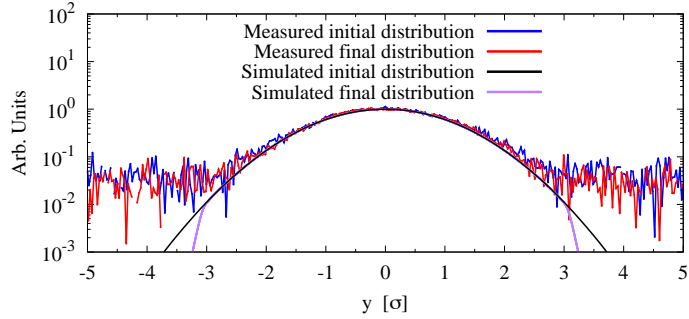}} 
\caption{Example of transverse beam profiles after blow up at the beginning and at the end of the loss measurements shown in Fig.~\ref{summary_plot} (c). The initial and final Gaussians agree very well. The initial and final beam distributions from the numerical simulations for the corresponding case are also shown.}
\label{transversedist}
\end{figure}

{ Figure~\ref{transversedist} shows the measured transverse profiles of the two beams after the emittance blow up at the beginning and at the end of the loss measurement reported in Fig.~\ref{summary_plot}~\mbox{(c)} (note that the following considerations hold true for all experimental configurations presented here). The profiles have been obtained by means of the synchrotron light monitor and the slight left-right asymmetry of the horizontal profile of Beam~2 is an artefact of the instrument and should be neglected~\cite{trad}. The values of the $\sigma$ of the two distributions are the same at the percent level and in general the two profiles match each other very well. The measurements are dominated by the noise whenever the transverse amplitude exceeds $\approx 2.5~\sigma$, while below this value, the transverse profiles prove to be Gaussian. The initial Gaussian distribution and the final one, as obtained from numerical simulations (see section~\ref{sec:anal}), are also shown. It is clearly seen that the diffusion mechanism is acting on the initial distribution by changing only the tails beyond $\approx 2.5~\sigma$. The typical losses observed are at the level of $1-2$\% of the bunch intensity, which agrees with the tail content of Gaussian beyond $\approx 2.5~\sigma$. Therefore, these observations suggest that a Gaussian initial distribution is an appropriate choice, although the synchrotron radiation monitor does not provide any direct quantitative measurement of the actual tails of the beam distribution.}

Note that in the rest of the paper the configuration in which all IR correctors are powered will be indicated as `with correctors', while that with the dodecapolar corrector only as `no correctors'.
\section{Modelling the experimental results with a diffusion equation} \label{sec:anal}
The experimental results presented in section~\ref{sec:exp} have been analysed by means of a 1D FP equation~\eqref{fokker2} with a Nekhoroshev-like form of the diffusion coefficient as in Eq.~\eqref{diffnek} { (parenthetically, the data from the measurement campaign performed at injection energy in the LHC have been re-analysed using the diffusive approach and the discussion of the results can be found in Ref.~\cite{arcidosso})}. { The model needs to find an efficient method to determine the three parameters. The first step is to constrain the model to agree with the measured intensity curve at the end of the experimental time window and this fixes $\varepsilon$. As a second step, the FP equation has been used to reproduce the various experimental cases determining the values of $\kappa$ and $I_\ast$ by minimising the $L^2$ norm of the difference between the solution of the FP equation with an absorbing boundary condition at $I=I_{\rm abs}$ and the corresponding measured intensity curve. This provided pairs of values for $\kappa_j, {I_\ast}_j$ for each experimental configuration. The main observation is that $\kappa$ is only very mildly depending on the configuration, which is in agreement with the fact that it should be linked with the number of degrees of freedom of the system under consideration. Therefore, the average of $\kappa_j$ has been used as an estimate of $\kappa$ for all data sets. The third and last step has been the computation of the solution of the FP equation for the various configurations by using the only remaining free parameter $I_\ast$ to minimise the $L^2$ as done for the second step.} It is worth pointing out that to remove the noise affecting the beam-intensity measurement, which is visible in the two rightmost plots of Fig.~\ref{summary_plot}, the intensity data have been filtered by a 50-data moving average. 

The procedure described above aims to point out the functional form of the beam losses produced by the Nekhoroshev's diffusion coefficient as we tried to avoid using the model parameters to obtain the best agreement with each individual measurement, but rather to find a global agreement between the numerical solutions and the measurement results. 

Since the action variable $I$ represents the non-linear invariant of the system, we have chosen as initial condition an exponential distribution
\begin{equation}
\rho_0(I)=\sigma^{-2}\, \exp\left (-\frac{I}{\sigma^2}\right )
\label{initial}
\end{equation}
where $\sigma^2$ stands for the measured beam emittance, in order to reproduce the measured beam profile { as shown in Fig.~\ref{transversedist}}. Moreover, by scaling the action variable $I\to I/\sigma^2$ we can set $\sigma=1$ in the simulations without affecting the beam-loss rate. Finally, the absorbing boundary condition $I_{\rm abs}$ is computed from the position of the collimator expressed in units of beam emittance and considering the physical plane where one expects the beam diffusion to be more relevant.

In Table~\ref{summary_table} we report the model's parameters obtained by applying the procedure described above, i.e. from the numerical evaluation of the relative intensity losses at the absorbing barrier with FP~\eqref{fokker2}. It is worth noting that for Beam~2 the case with H blow up and `with correctors' features no appreciable beam losses and therefore, no attempt to derive model's parameters has been made. 

\begin{table}
\centering
\caption{Summary of the model's parameters obtained with the numerical simulations of the measured beam losses, { using the approach described in the main text}. In the case of Beam~2 with horizontal blow up, the losses for the configuration `with correctors' are not high enough to attempt any meaningful modelling. The plane where the absorbing boundary is set is specified in parenthesis for the case with H-V blow up. There, the boundary condition is set in the plane and to the value corresponding to the minimum amplitude between the boundary conditions in the horizontal and vertical planes. { The $L^2$ norm is also given, which is to be considered relative to the total beam losses measured for each configuration.}}
\label{summary_table}
\begin{tabular}{lccccc}
\hline 
\multicolumn{2}{c}{} & Beam~1 & \multicolumn{3}{c}{Beam~2} \\
Configuration & Parameter & H-V  & H-V & V & H               \\ 
\hline 
\multirow{4}{15truemm}{No corr.}& $\kappa$ &$0.33$ & $0.33$ & $0.33$ & $0.33$\\
&  $\epsilon \,[10^{-4}]$    &$4.7$     & $10.0$    & $1.5$ & $0.9 $\\
&  $I_\ast \,[\sigma^2]$     &$8.0$     & $14.0$    & $21.5$& $21.0$\\
&  $I_{\rm abs} \,[\sigma^2]$&$7.8$ (H) & $8.0$ (H) & $8.0$ & $7.1 $\\ 
& $L^2$ norm                 &$0.7$     & $1.2$     & $4.8$ & $3.0$\\
& $(10^{-2})$                &$ $       & $ $       & $ $   & $ $\\\hline 
\multirow{4}{15truemm}{With corr.}& $\kappa$ &$0.33$ & $0.33$ & $0.33$ & -- \\
&  $\epsilon \,[10^{-4}]$    &$8.2$     & $17.0$    & $13$  & -- \\
&  $I_\ast \,[\sigma^2]$     &$9.5$     & $14.0$    & $21.5$& -- \\
&  $I_{\rm abs} \,[\sigma^2]$&$7.8$ (H) & $7.8$ (H) & $7.9$ & -- \\
& $L^2$ norm                 &$3.2$     & $1.2$     & $1.2$ & -- \\
& $(10^{-2})$                & $ $      & $ $       & $ $   & $ $ \\\hline 
\hline
\end{tabular}
\end{table}

{ The values of the $L^2$ norm for the final numerical results are also listed in Table~\ref{summary_table}. The norm provides a cumulative measure of the deviation between the measured and the simulated beam-loss curves and the values are relative to the total beam loss measured for each configuration. The order of magnitude is around few percent, which corresponds to about few $10^{-4}$ of the absolute intensity loss. Note that the precision with which the beam intensity is measured is below the percent level. Therefore, the overall agreement can be considered excellent.}

\begin{figure}[htb]
\centering
\begin{tabular}{@{}c@{}}
{\includegraphics[trim= 5mm 60mm 20mm 50mm, width=0.6\linewidth,clip=]{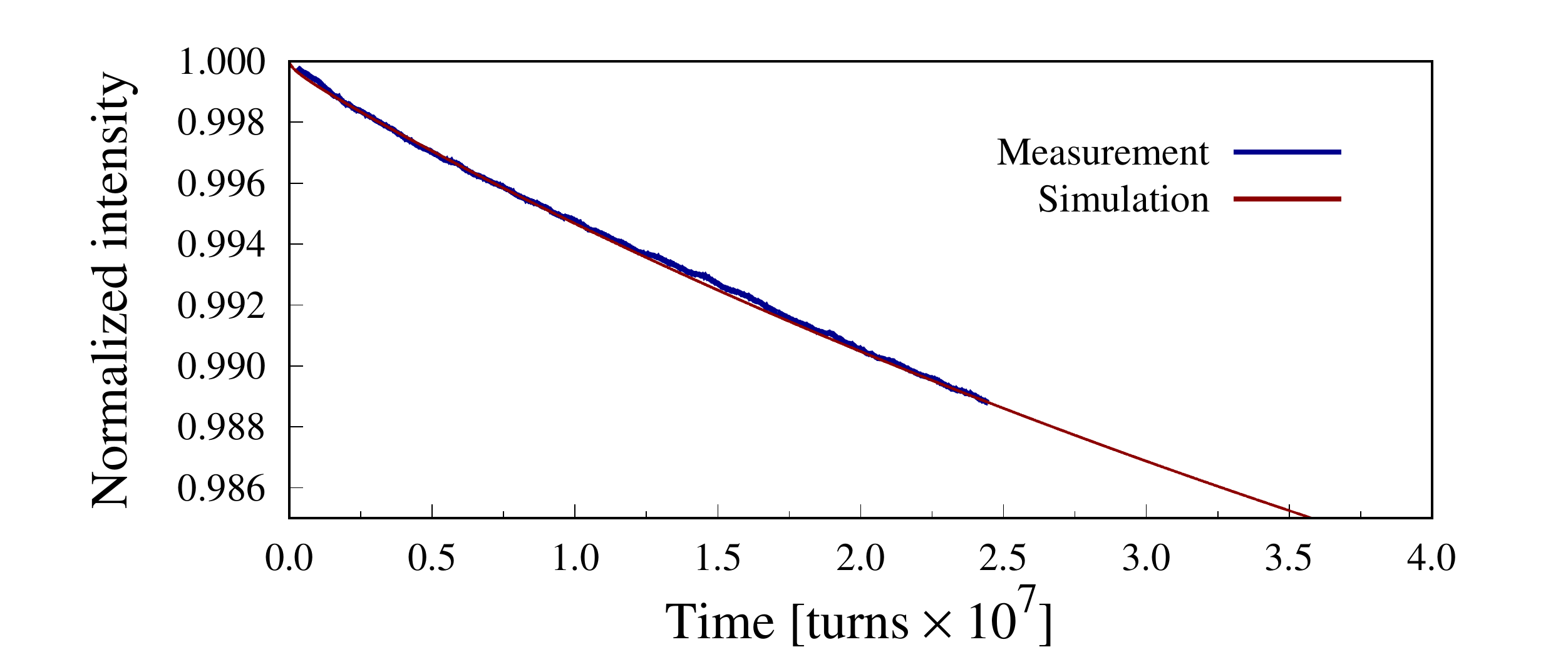}} \\
{\includegraphics[trim= 5mm 60mm 20mm 50mm, width=0.6\linewidth,clip=]{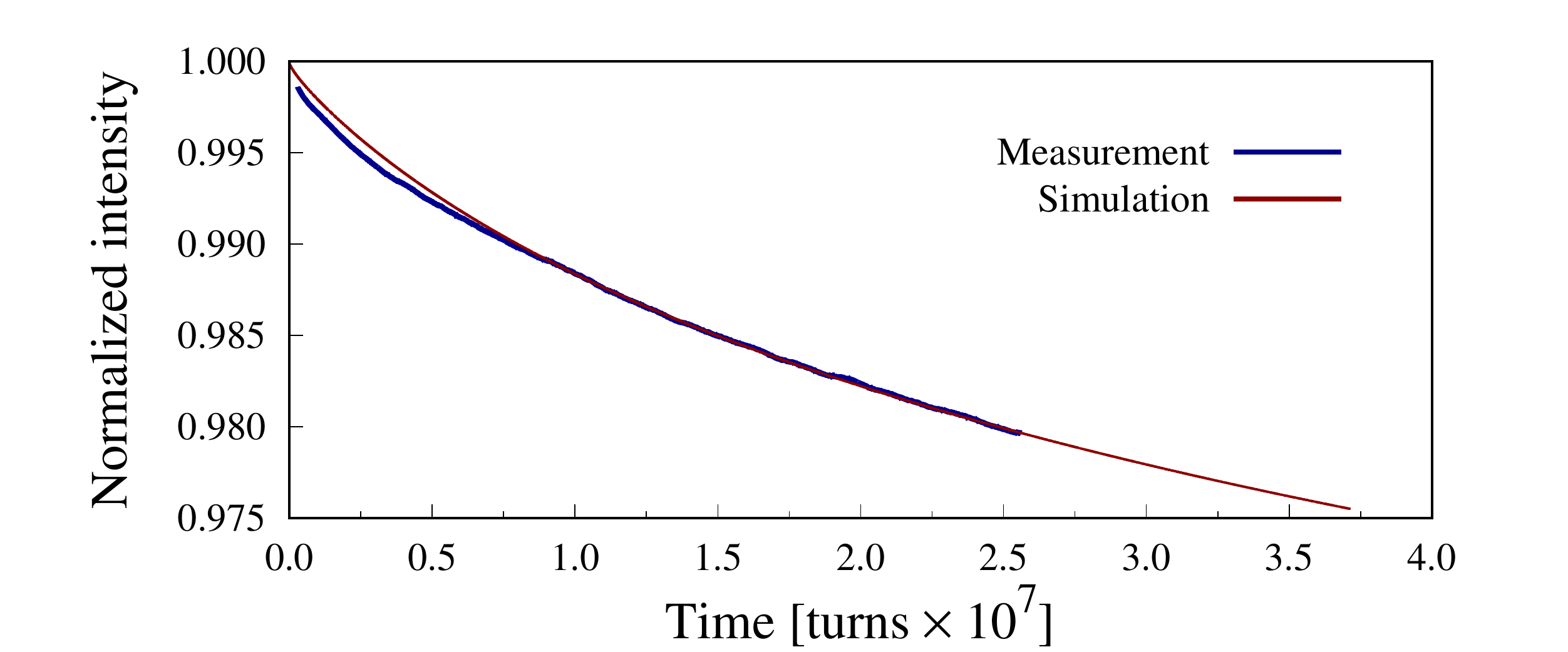}} 
\end{tabular}
\caption{{ Measured and simulated} intensity loss for the Beam~1 data set with H-V blow-up by using the 1D FP equation~(\ref{fokker2}) for configuration `no correctors' (upper) and `with correctors' (lower). The initial distribution is exponential and the values of the model parameters are those reported in Table~\ref{summary_table}.}
\label{beam1fig}
\end{figure}

Figure~\ref{beam1fig} shows the results of the numerical simulations together with the experimental data for the complete Beam~1 data set. 

\begin{figure}[htb]
\centering
\begin{tabular}{@{}c@{}}
{\includegraphics[trim= 5mm 60mm 20mm 50mm, width=0.6\linewidth,clip=]{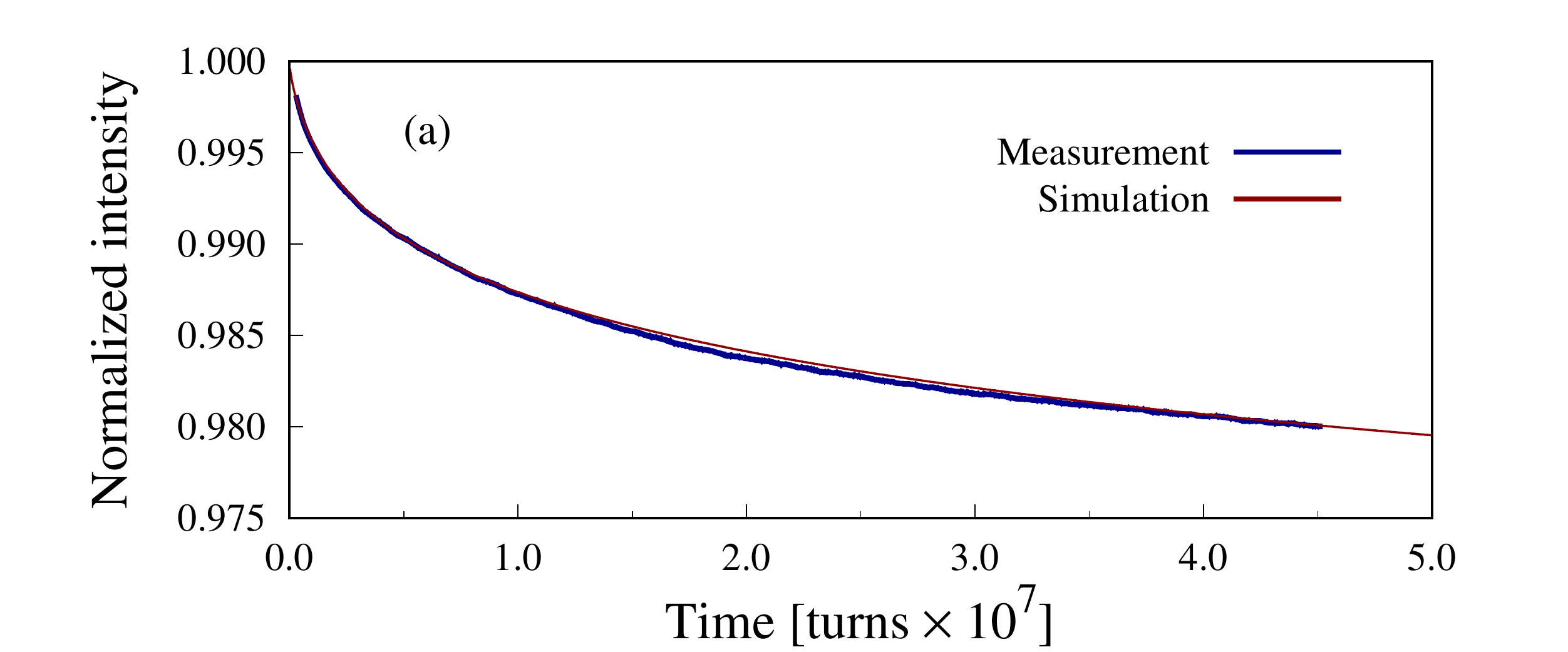}} \\
{\includegraphics[trim= 5mm 60mm 20mm 50mm, width=0.6\linewidth,clip=]{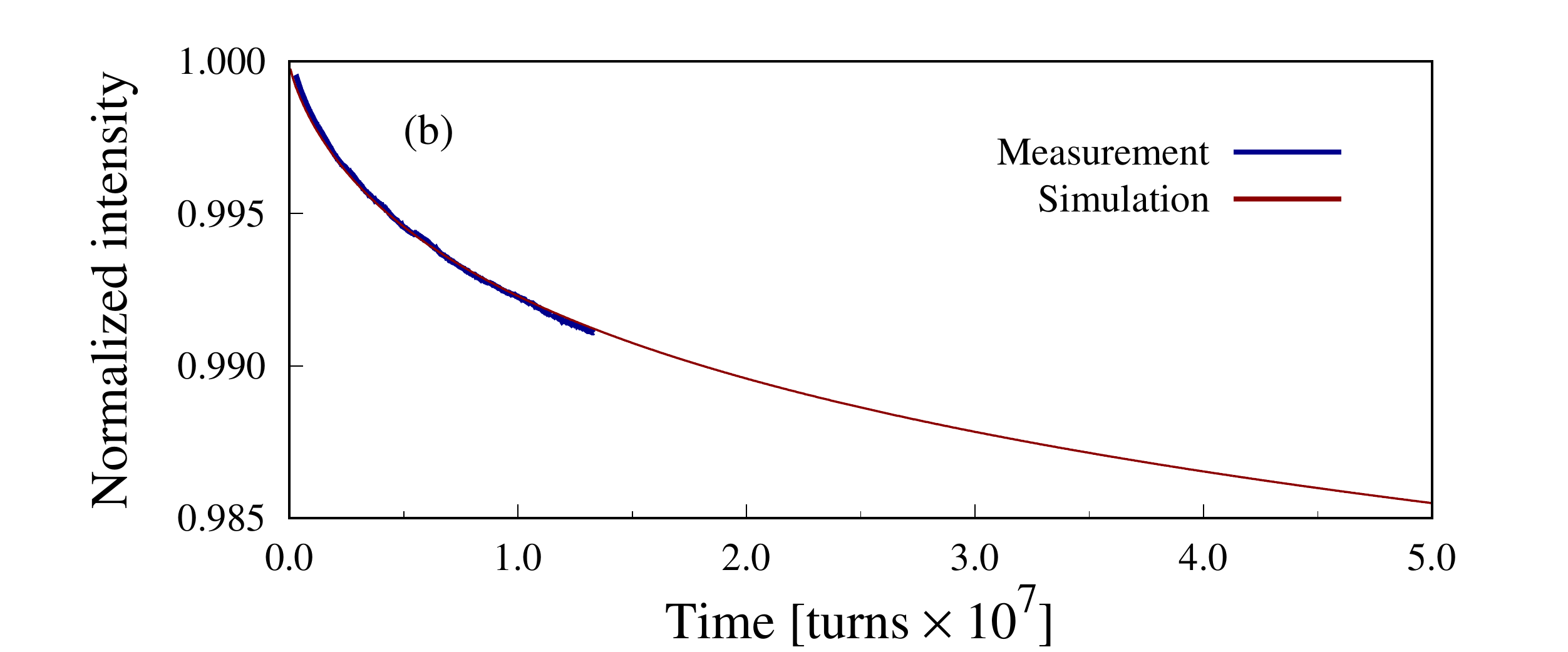}} \\
{\includegraphics[trim= 5mm 60mm 20mm 50mm, width=0.6\linewidth,clip=]{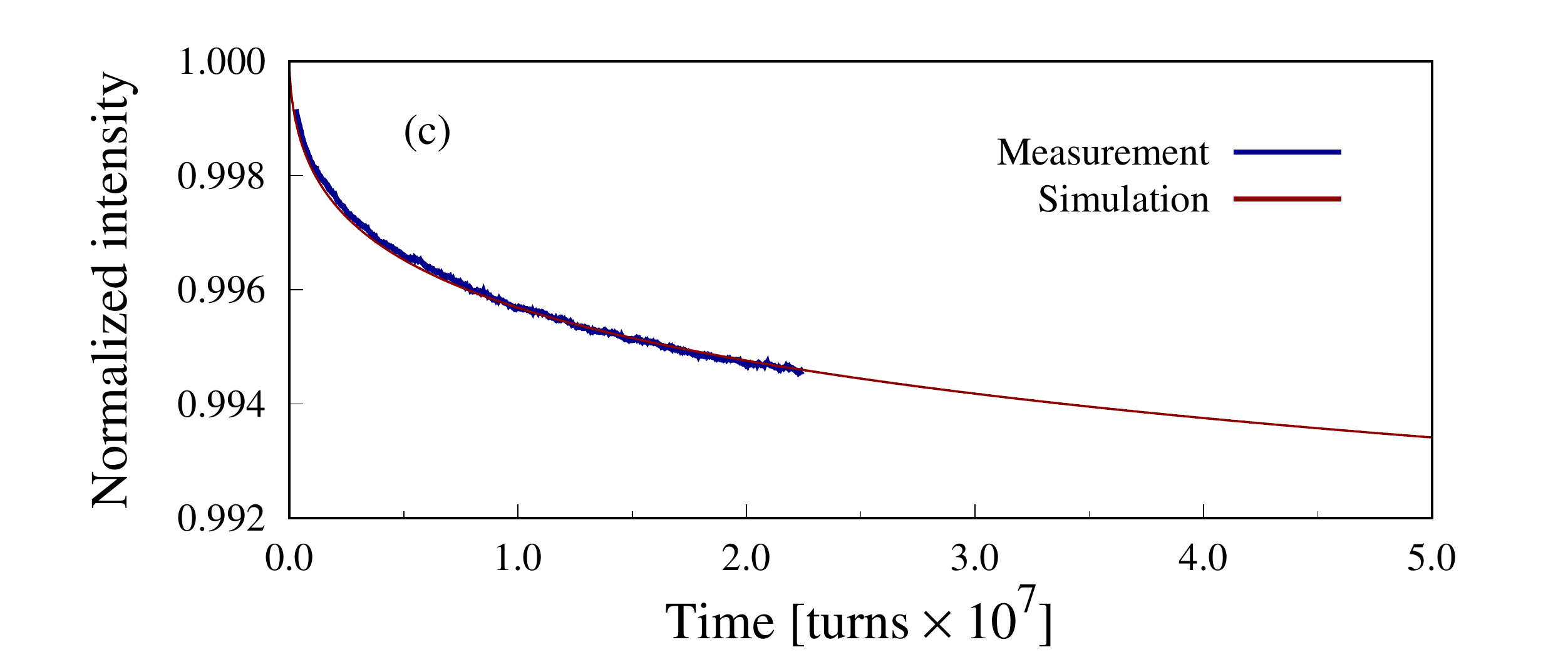}} \\
{\includegraphics[trim= 5mm 60mm 20mm 50mm, width=0.6\linewidth,clip=]{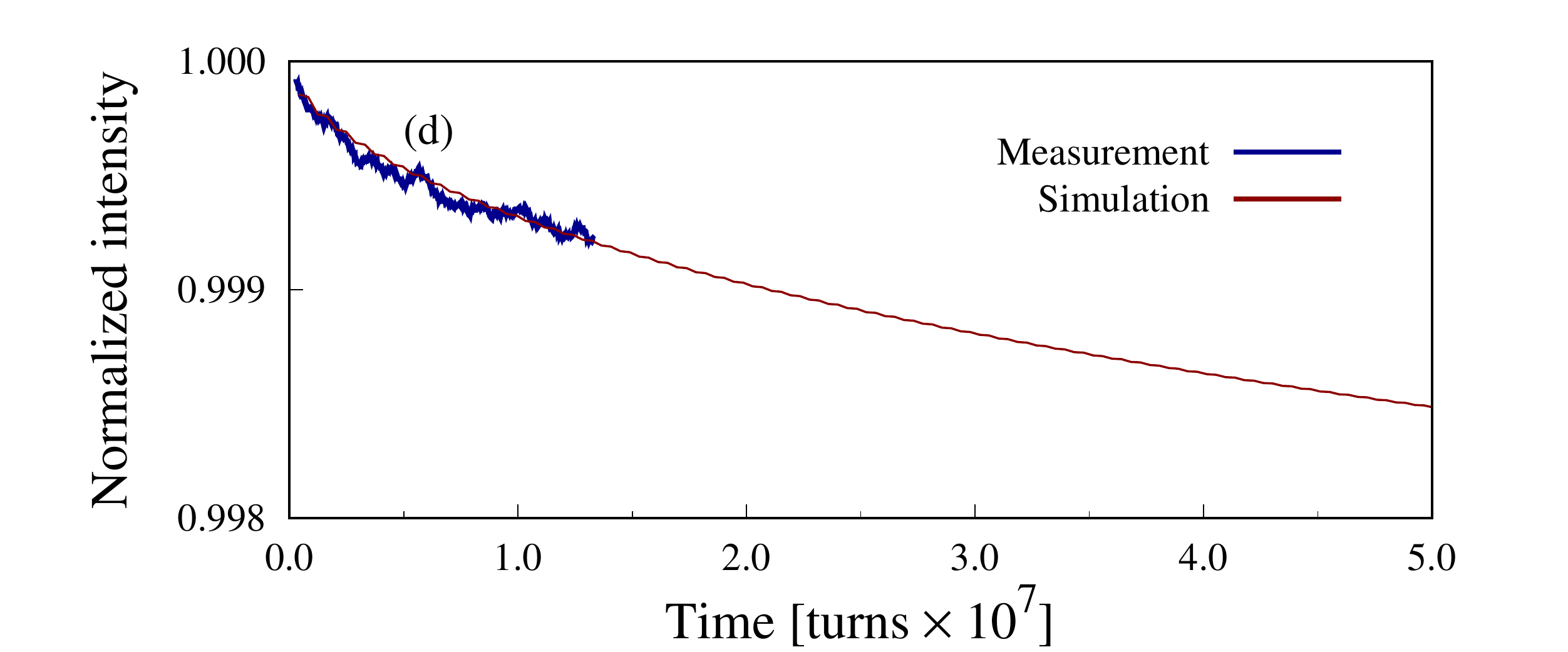}}
\end{tabular}
\caption{{ Measured and simulated} intensity loss for the Beam~2 data sets with H-V blow and `with correctors' (a); with H-V blow and `no correctors' (b); with V blow and `with correctors' (c); with V blow and `no correctors' (d). The initial distribution is exponential and the values of the model parameters are those reported in Table~\ref{summary_table}.}
\label{beam2figv_hv}
\end{figure}

The agreement between the measured data and the simulations results { is excellent as it is indicated by the values of the $L^2$ norm in Table~\ref{summary_table}.} Figure~\ref{beam2figv_hv} shows the results of our analysis for the Beam~2 data set, in which the case with H blow up has been discarded due to insufficient level of beam losses.

Also in this case the agreement between experimental observations and numerical simulations is striking. It is also worth noting that the time span of the various data sets covers a rather wide range of turn numbers and the agreement does not depend on the duration of the measurements. 

According to the physical interpretation of the diffusive model parameters, the exponent $\kappa$ plays a fundamental role in determining the shape of the beam-loss curve. The second model's parameter $I_\ast$ defines a transition threshold in the action space from a fast to a slow diffusion and it changes the shape of the curve when its value is comparable with the position of the absorbing barrier. 

\begin{figure}[htb]
\centering
\begin{tabular}{@{}c@{}}
{\includegraphics[trim= 5mm 60mm 20mm 50mm, width=0.6\linewidth,clip=]{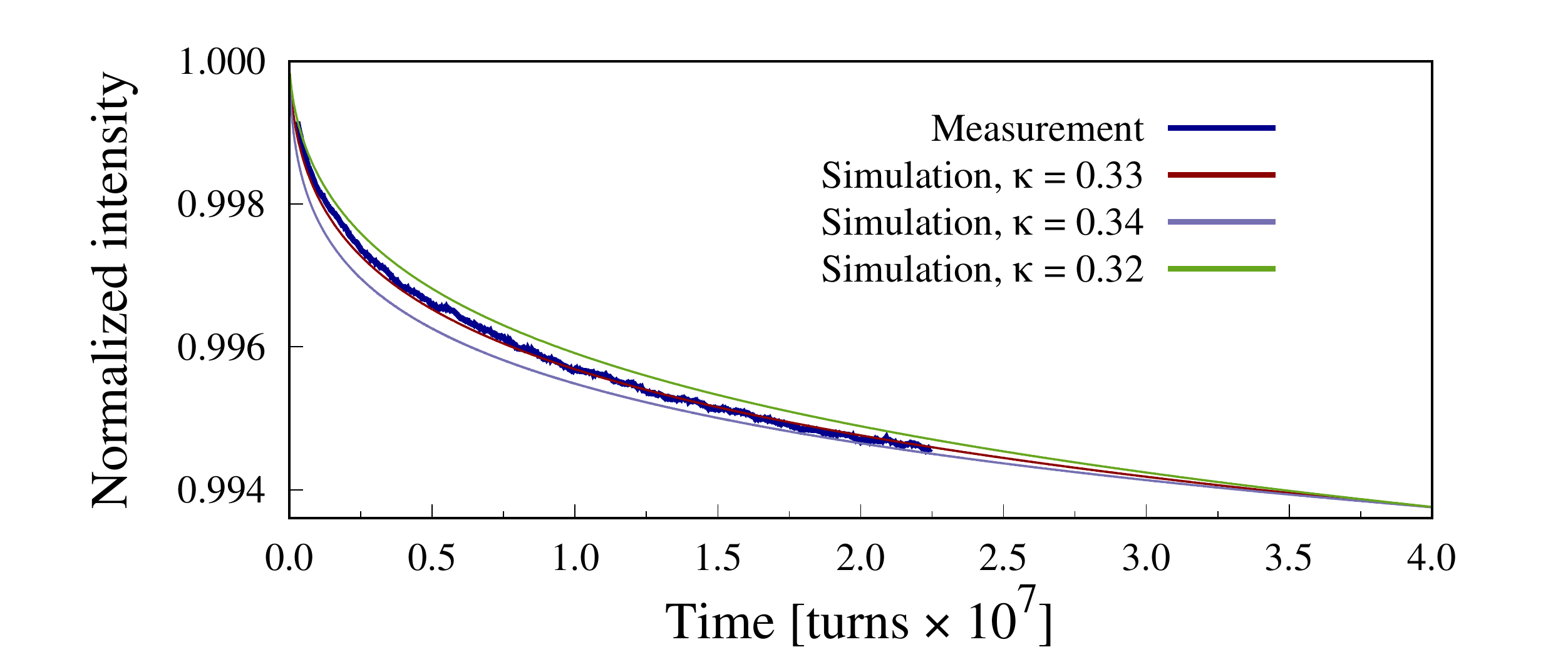}} \\ 
{\includegraphics[trim= 5mm 60mm 20mm 50mm, width=0.6\linewidth,clip=]{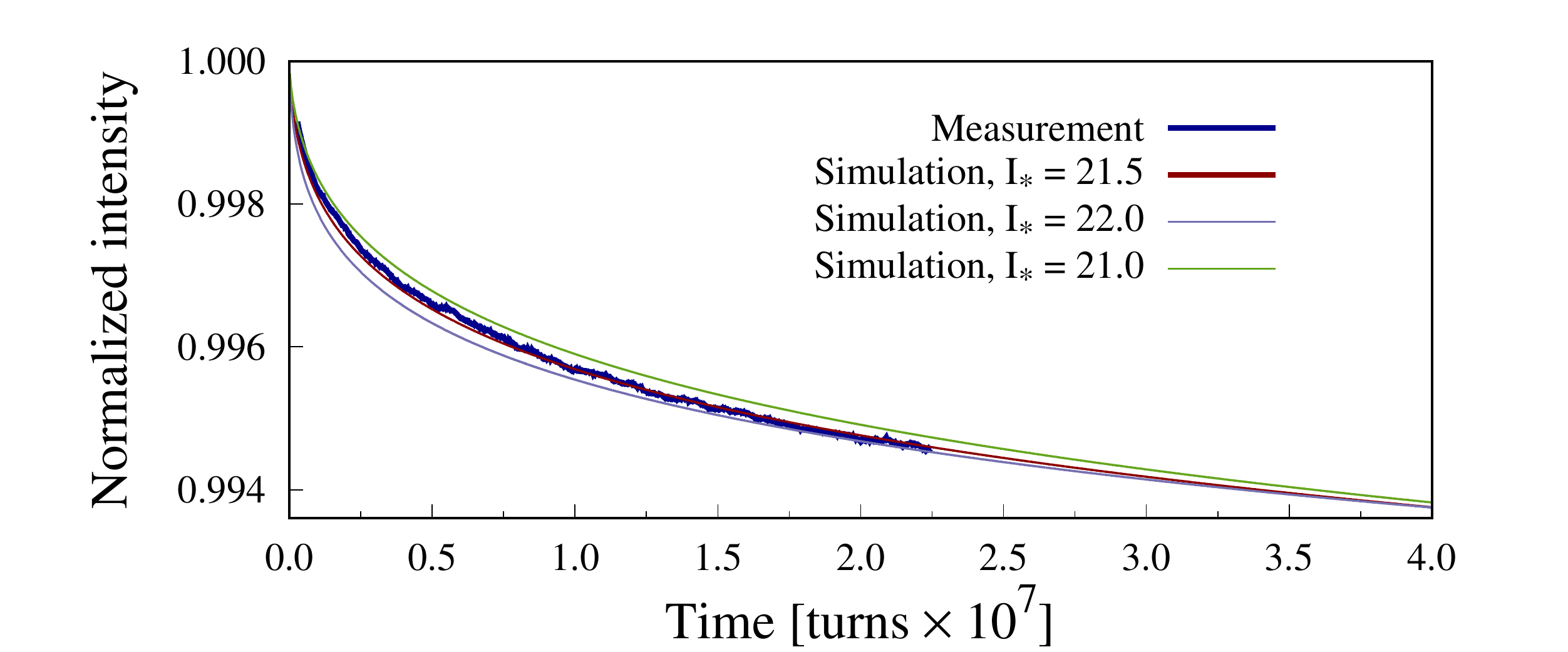}} \\
{\includegraphics[trim= 5mm 60mm 20mm 50mm, width=0.6\linewidth,clip=]{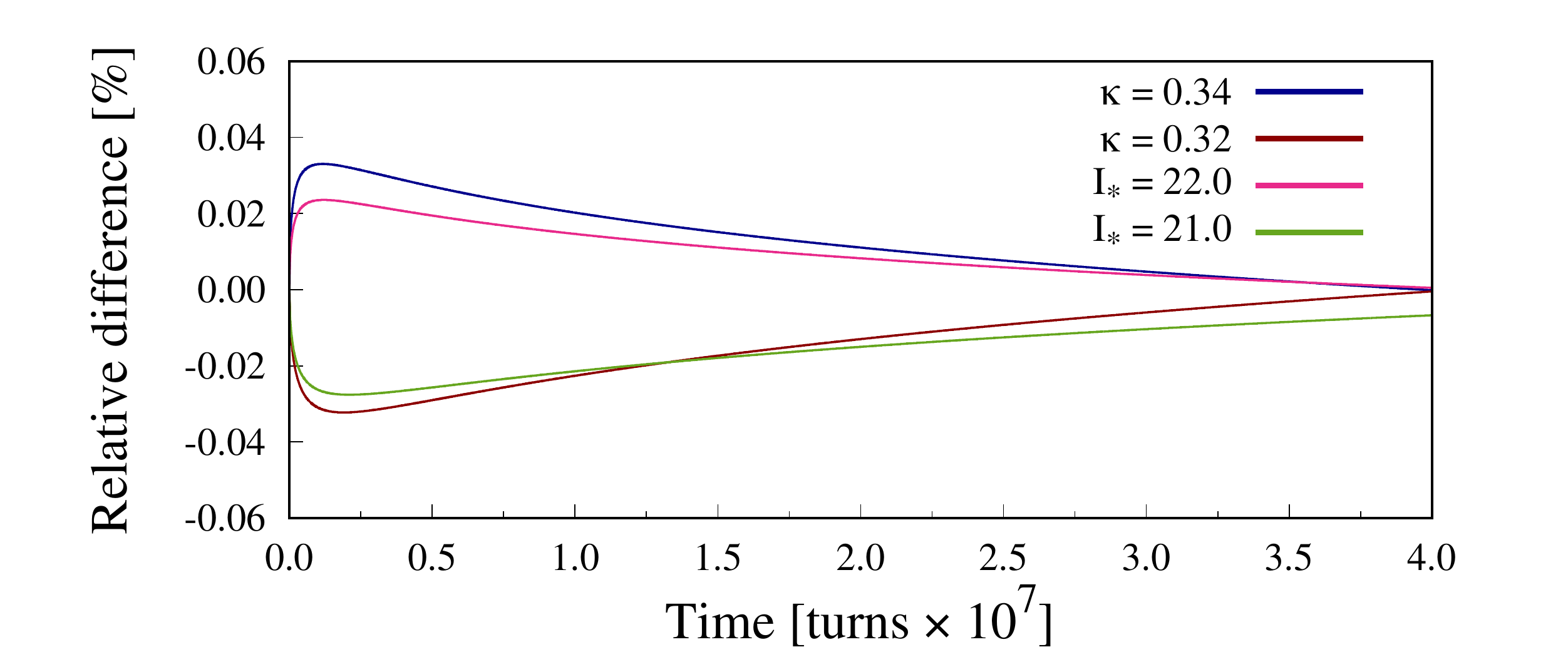}} 
\end{tabular}
\caption{Loss curves for Beam~2 with V blow up for configuration `with correctors' showing also the results for models in which the parameter $\kappa$ (upper) or $I_\ast$ (lower) is varied around the optimal value. In both cases the parameter $\varepsilon$ is adapted to fix the total losses, so that all  curves intersect at the end of the time interval. { In the bottom plot the relative difference between the curve reproducing the experimental data and those with varied model parameters is shown. The variation of $\kappa$ produces the largest change in the loss curve.}}
\label{sensitivityfig}
\end{figure}

To illustrate the sensitivity of the simulated beam intensity to the values of $\kappa$ and $I_\ast$ in  Fig.~\ref{sensitivityfig} we compare the beam-loss curves computed with the diffusion model when the two parameters are varied, one at a time, with respect to the optimal value { top and center plots. In the bottom plot the relative difference between the curve reproducing the experimental data and those with varied model parameters is shown.}

The value of $\kappa$, varied by few percent, influences substantially the shape of the initial part of the interpolating curve, right after the initial fast transient beam losses. This observation supports the assumption that the constancy of the exponent $\kappa$ for the different considered cases can be attributed to an intrinsic property of the observed diffusion process. 

In the perturbation theory, the $I_\ast$ parameter is interpreted as a global scaling for the perturbative series related to the nature of the non-linear terms present in the system, although it is not directly linked to their magnitude. The effect of the value of $I_\ast$ on the shape of the beam-loss curve depends on the ratio $I_{\rm abs}/I_\ast$, which provides the position of the absorbing barrier: a greater value of $I_\ast$ reduces the beam halo and consequently the beam losses at the position of the absorbing barrier, whereas the opposite effect occurs for lower values of $I_\ast$. 

In summary, Fig.~\ref{sensitivityfig} indicates that the proposed approach is sensitive to changes in $\kappa$ and $I_\ast$ at the level of few percent, which { provides a very strong support to the robustness of the proposed model against variation of its parameters.} In turns, this means that the differences between the values obtained by the numerical simulations could reflect actual differences in the dynamics occurring in the weak chaotic regions where the diffusion phenomena take place. 
\section{Symplectic-tracking checks of the experimental obervations} \label{sec:tracking}
{ The analysis presented in this paper does not rely on anything else than the numerical solution of the FP equation. Nevertheless, some tracking simulations have been performed to assess the choice of the boundary conditions and the plane of losses as well as some of the assumptions needed for the diffusive approach to be a valid option.}

The ring model is the most accurate description of the LHC lattice including the measured field errors (see~\cite{DAasbuilt} for more detail) together with the operational configuration of the various correction circuits. The numerical protocol used envisages the generation of sixty realisations of the magnetic errors to take into account the measurement uncertainties, moreover, a polar grid of initial conditions in $x-y$ space is defined and its evolution is computed for up to $10^6$ turns. The polar grid of initial conditions is obtained by dividing the first quadrant of the $x-y$ space in $59$ angles and along each direction $30$ initial conditions are uniformly distributed over intervals of $2\sigma$. 

\begin{figure}[htb]
\centering
\begin{tabular}{@{}c@{}@{}c@{}}
{\includegraphics[trim = 20mm 20mm 60mm 20mm, width=0.35\linewidth,clip=]{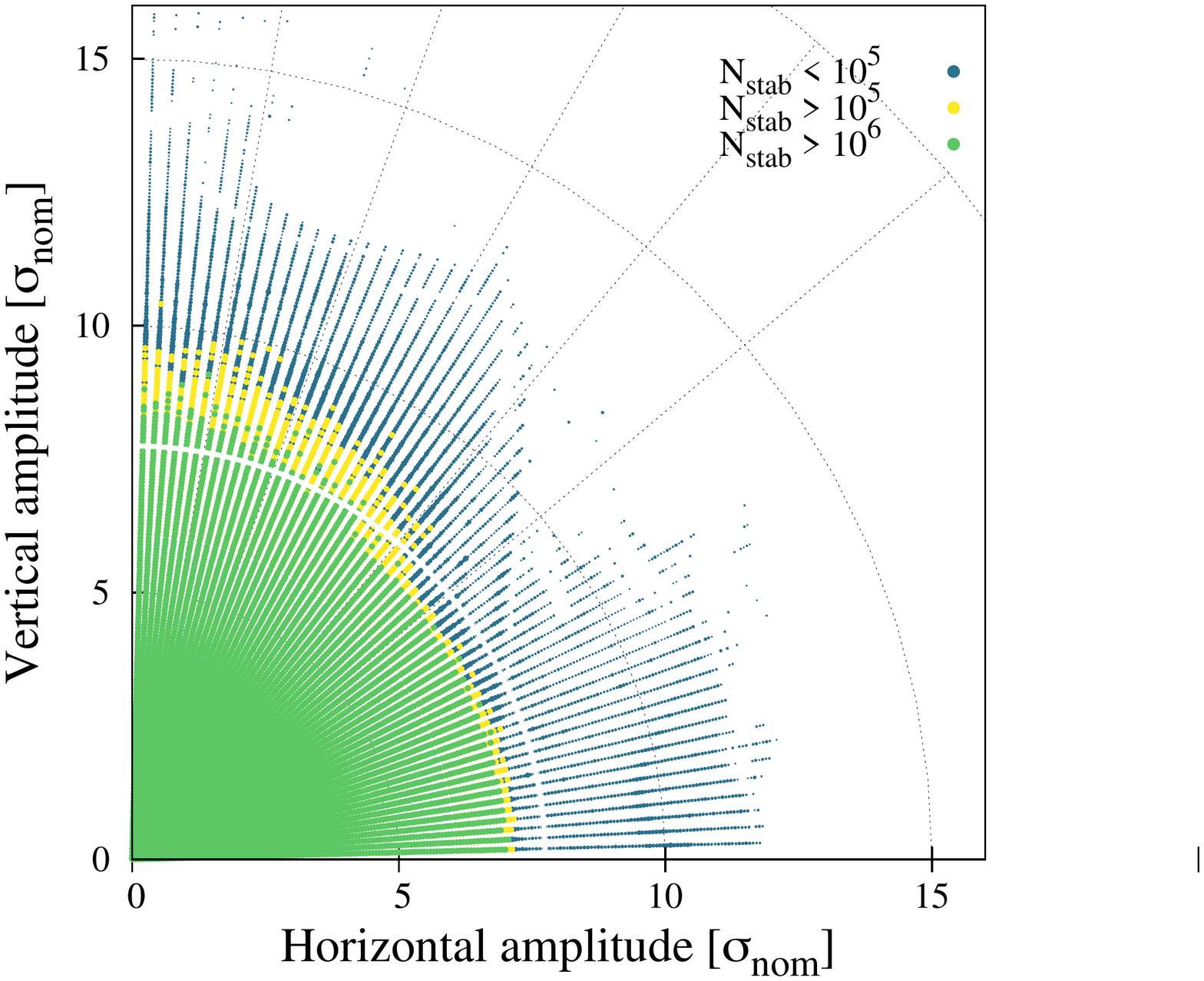}} &
{\includegraphics[trim = 20mm 20mm 60mm 20mm, width=0.35\linewidth,clip=]{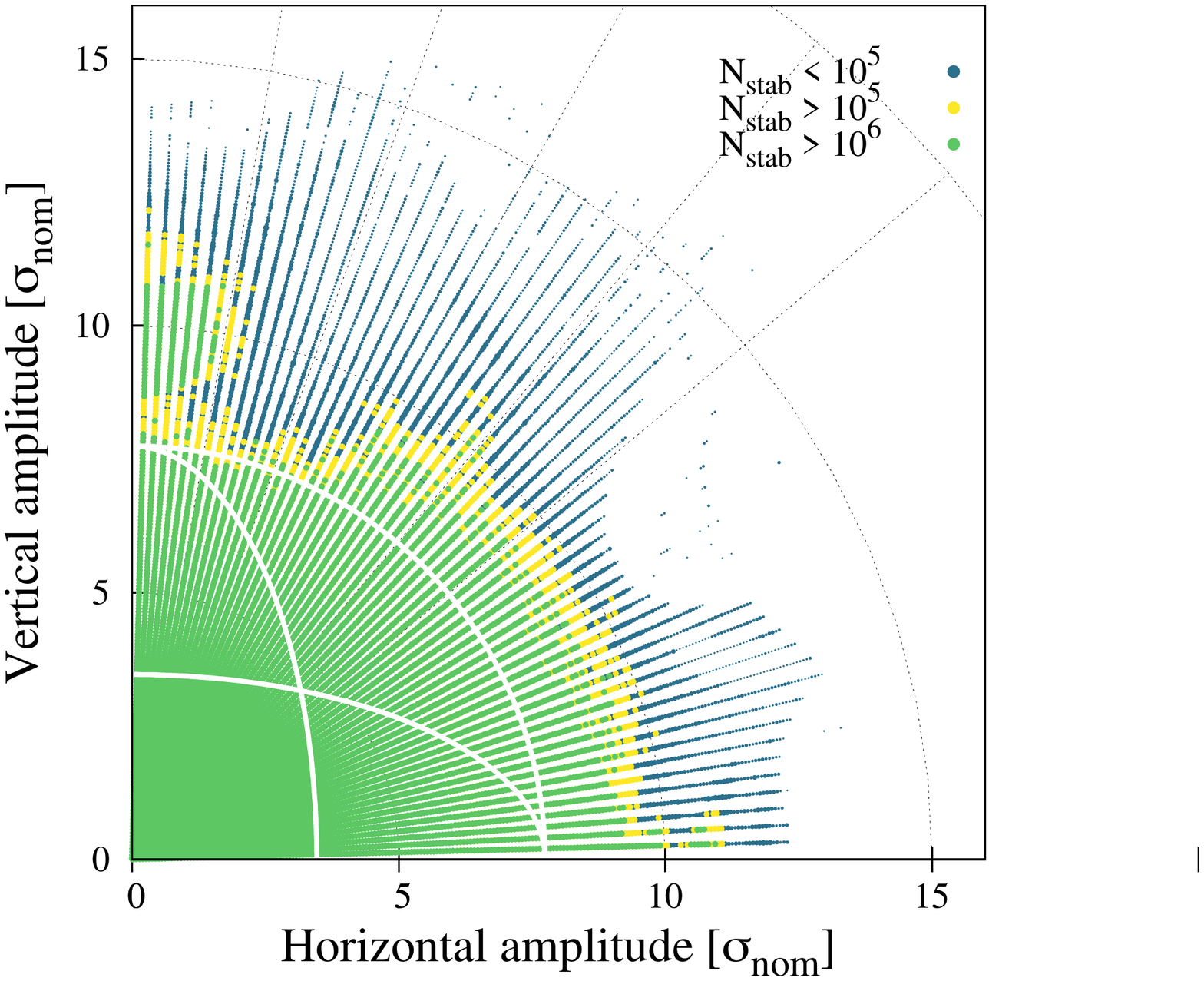}} \\
{\includegraphics[trim = 20mm 20mm 60mm 20mm, width=0.35\linewidth,clip=]{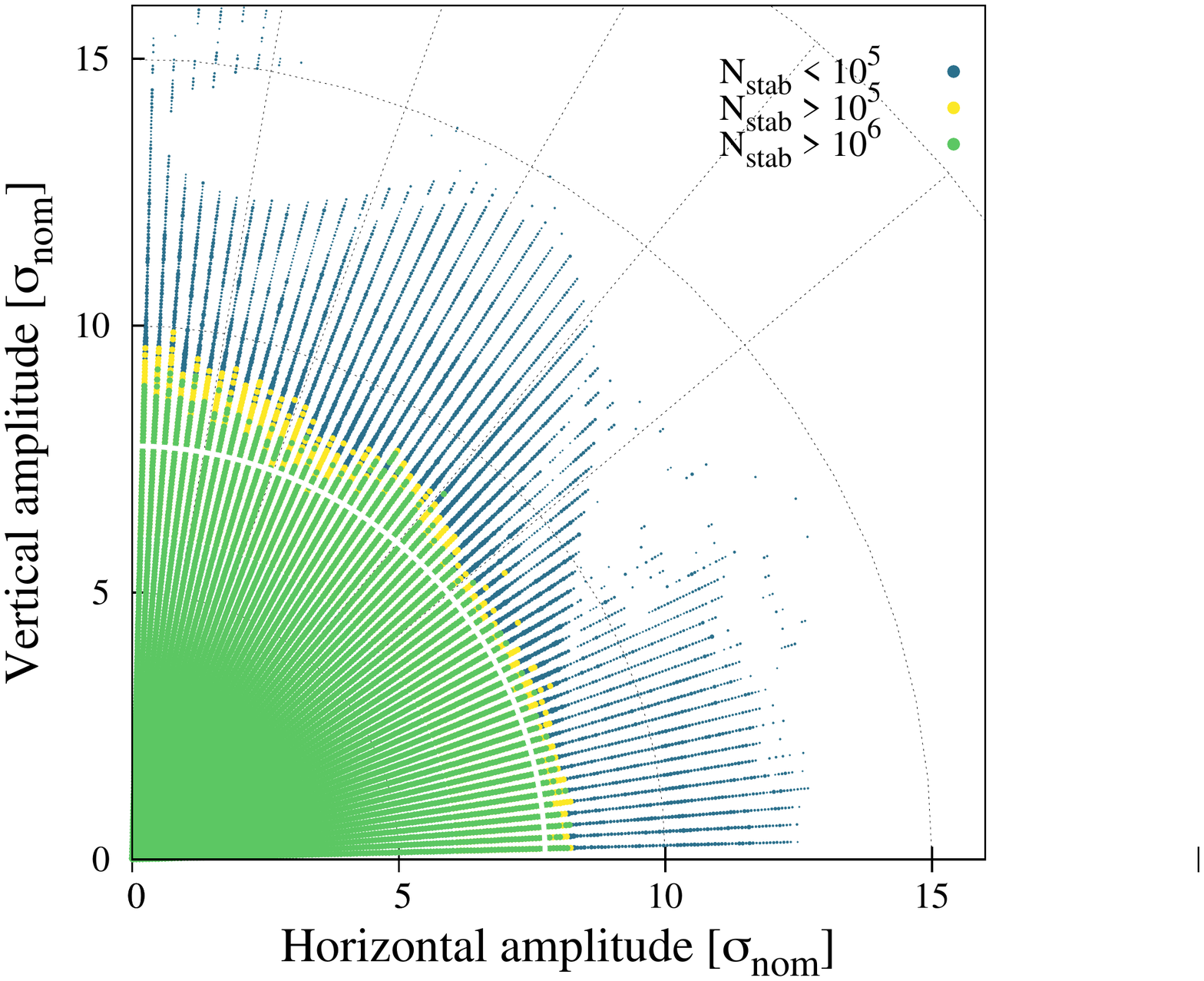}} &
{\includegraphics[trim = 20mm 20mm 60mm 20mm, width=0.35\linewidth,clip=]{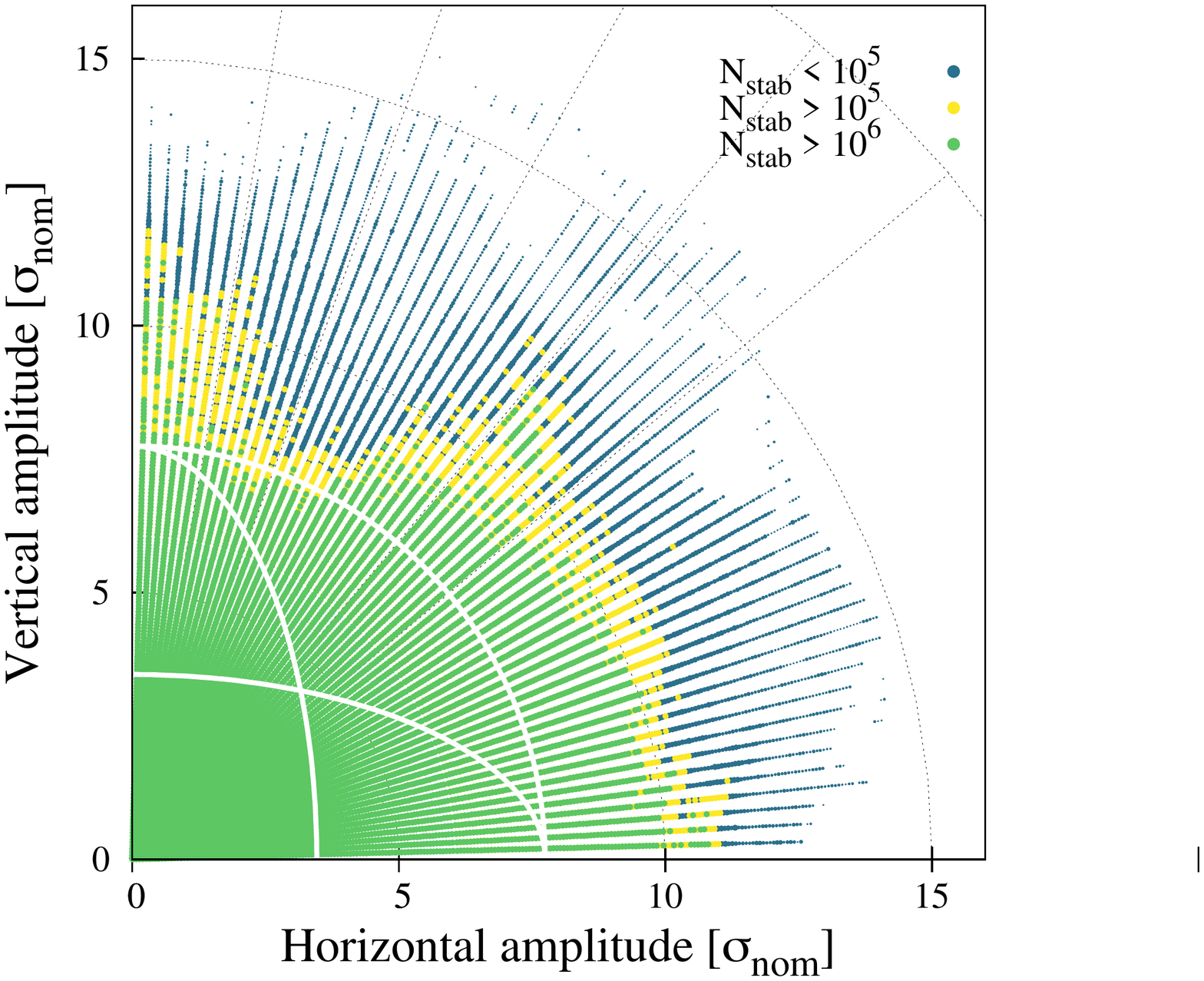}} \\
\end{tabular}
\caption{Plots of the stable region in $x-y$ space for Beam~1 (upper row) and Beam~2 (lower row) for the first realisation of the magnetic errors. The configuration `with correctors' is shown in the left column, while that with `no correctors' in the right one. The various colours indicate different stability time $N_{\rm stab}$ and initial conditions that are not stable for at least $10^5$ turns are represented by a marker whose size is proportional to the stability time. The white lines represent the $3\sigma$ level lines of the beam distribution for the three types of blow up, namely H, V, and H-V.}
\label{da_plot}
\end{figure}

The evolution of the initial conditions through the LHC lattice is computed using the SixTrack code~\cite{sixtrack}, which implements a second-order symplectic integration method. The loss time, i.e. the time an orbit associated with a given initial condition reaches a pre-defined amplitude, is recorded and associated to each initial condition. The outcome of these simulations is shown in Fig.~\ref{da_plot}, where the stable region is shown for Beam~1 (upper row) and Beam~2 (lower row) and for each of the two configurations used in the experiment (`with correctors' in the left column, `no correctors' in the right one) for the first realisation of the magnetic errors.

The different colours are used to identify various stability times $N_{\rm stab}$, i.e. dark-blue markers indicate particles with $N_{\rm stab} < 10^5$ and the marker size is proportional to the value of $N_{\rm stab}$. Yellow markers indicate a region for which $N_{\rm stab} > 10^5$, while for green markers $N_{\rm stab} > 10^6$. The shrinking of the extent of the stable region for increasing values of $N_{\rm stab}$ is clearly visible. Moreover, the border of stability is almost circular for Beam~1, whereas it is much more irregular for Beam~2. 

Figure~\ref{da_plot} shows also three white curves: they represent the $3\sigma$ level lines of the beam distribution for the three types of blow up applied during the experiment, namely H, V, or H-V. For Beam~1, the beam distribution is rather close to the stability boundary and sizeable losses are to be expected. For Beam~2, it is worth noting that the irregular shape of the stability border implies that the $3\sigma$ edge of the beam distribution is relatively far from the border itself, which is in agreement with the low beam losses measured for the H blow up case. Both the H-V and V blow up cases feature the edge of the beam distribution close to the stability border in the vicinity of the $y$ axis. This explains qualitatively the higher losses observed for these cases and, of course, also the fact that the case with H-V blow up generates even higher beam losses.

\begin{figure}[htb]
\centering
\begin{tabular}{@{}c@{}@{}c@{}}
{\includegraphics[trim = 20mm 20mm 60mm 20mm, width=0.35\linewidth,clip=]{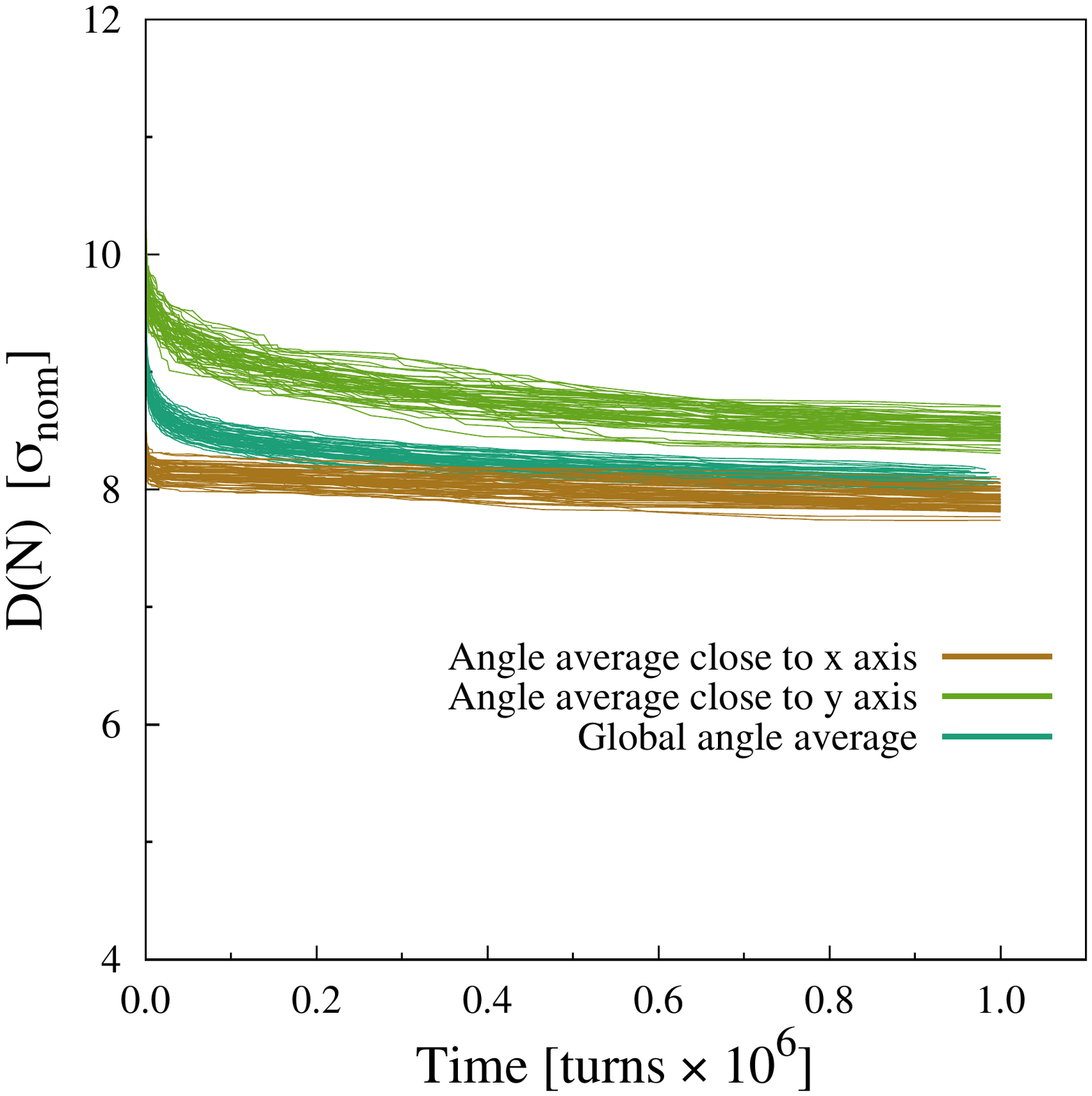}} &
{\includegraphics[trim = 20mm 20mm 60mm 20mm, width=0.35\linewidth,clip=]{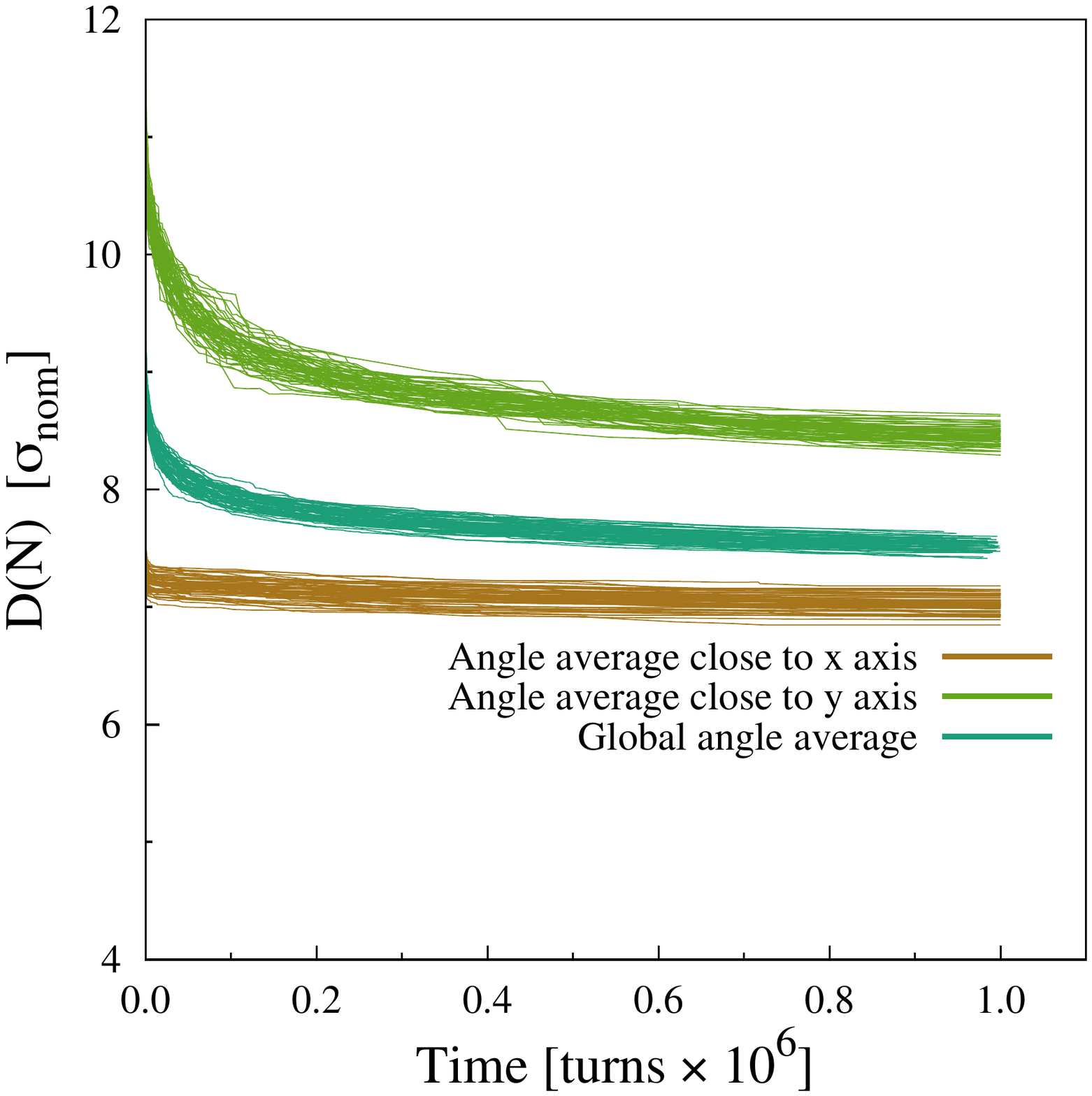}} \\
{\includegraphics[trim = 20mm 20mm 60mm 20mm, width=0.35\linewidth,clip=]{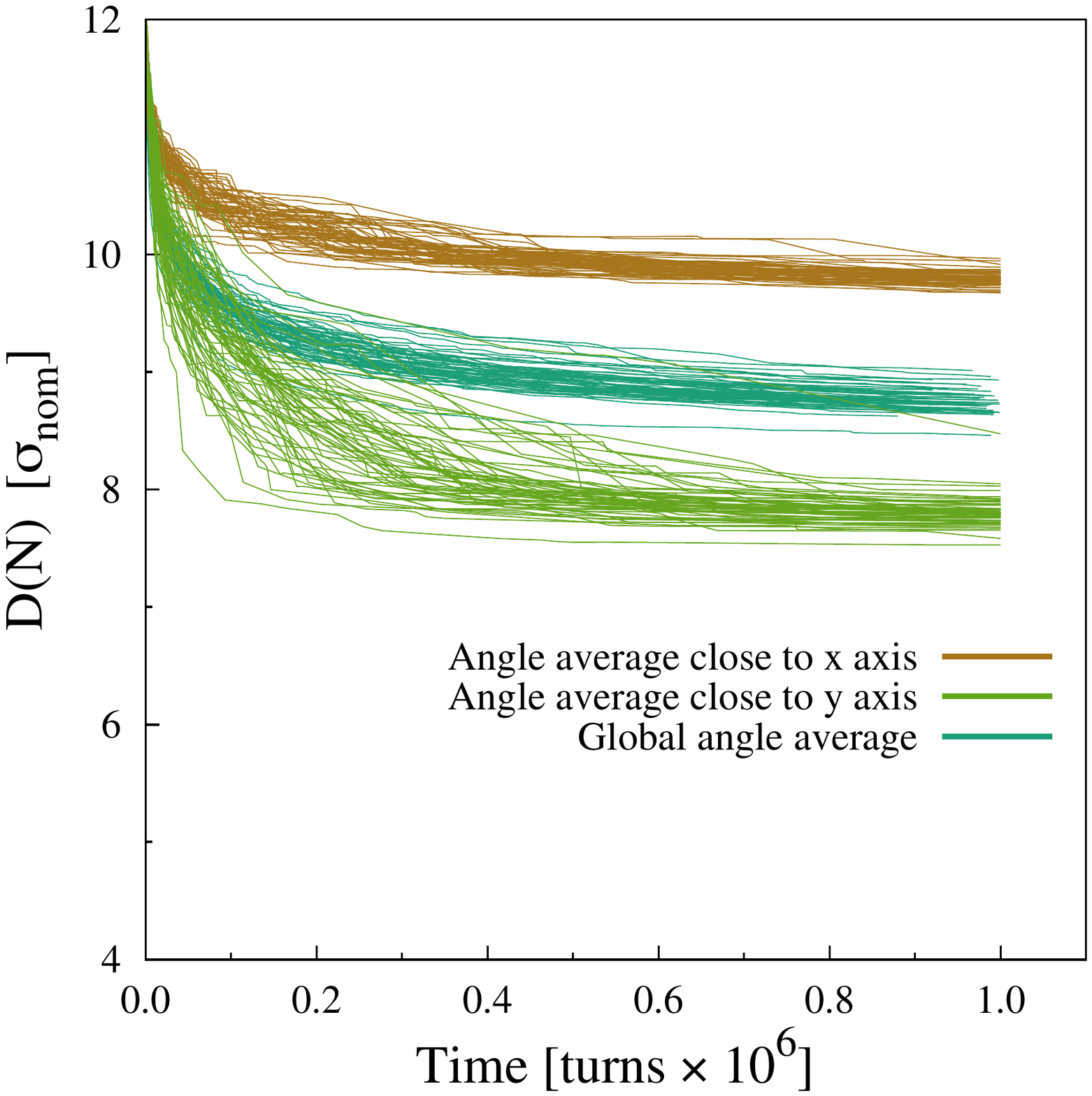}} &
{\includegraphics[trim = 20mm 20mm 60mm 20mm, width=0.35\linewidth,clip=]{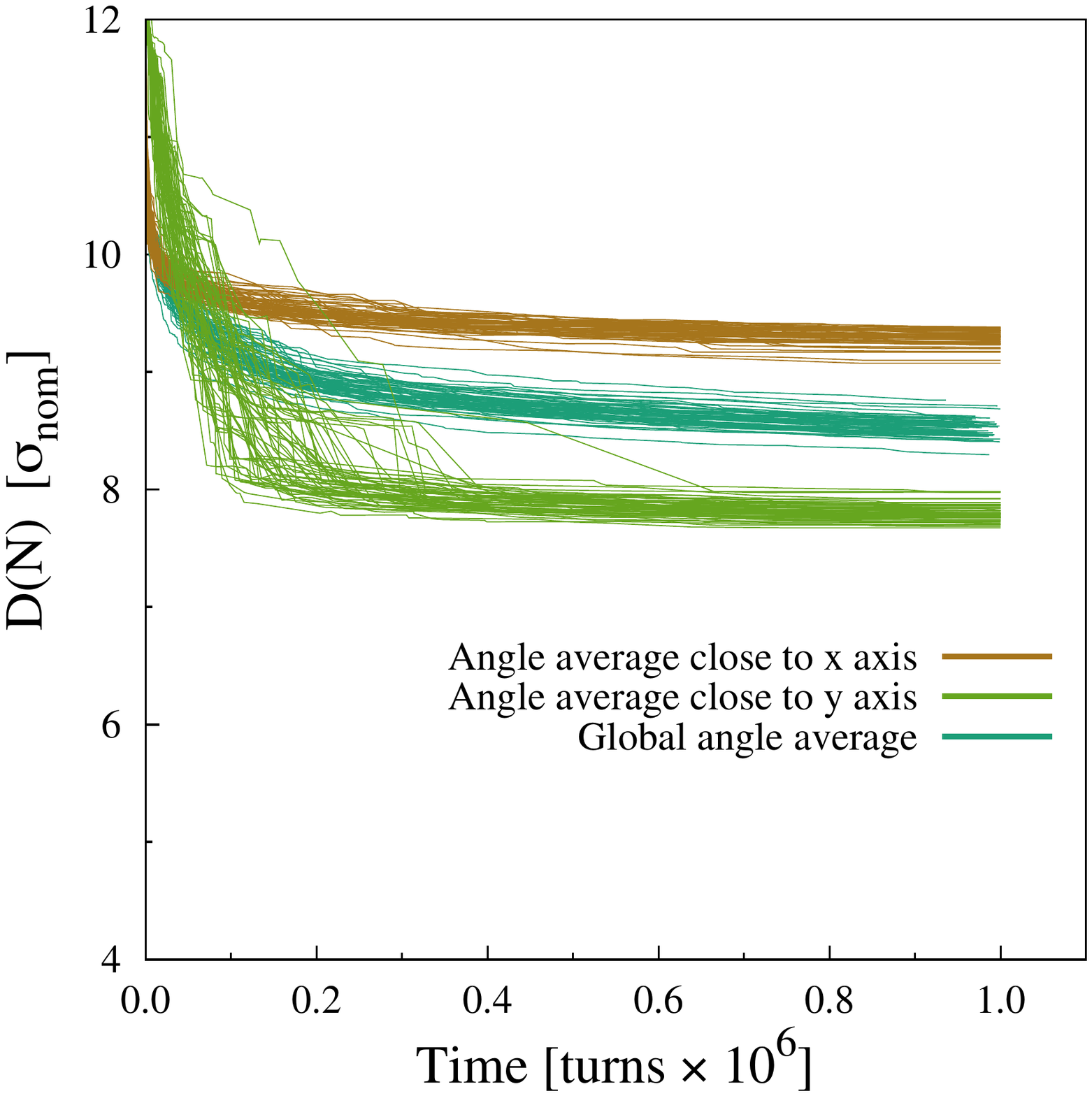}} \\
\end{tabular}
\caption{Plots of the DA evolution with turn number for Beam~1 (upper row) and Beam~2 (lower row). The configuration `with correctors' is shown in the left column, while that with `no correctors' in the right one. The results for all sixty realisations of the LHC lattice are reported. The three sets of curves refer to the DA averaged over all angles (see Eq.~\eqref{DAdef}) or averaged over $10$ angles close to the $x$ or $y$ axis, respectively.}
\label{da_vs_time}
\end{figure}

These plots, however, provide only a static information about the extent of the stable region of phase space. The time-dependence can be reconstructed by means of Eq.~\eqref{DAdef} and it is shown in Fig.~\ref{da_vs_time}, where the Beam~1 (upper row) and the Beam~2 (lower row) cases are shown. The configuration `with correctors' is reported in the left column, whereas that with `no correctors' in the right one. 

Each plot features three sets of curves: one representing the DA averaged over all angles in the $x-y$ space according to Eq.~\eqref{DAdef} { (hence representing a situation relevant for the H-V blow-up case)}; one representing the DA averaged over $10$ angles, only, in the vicinity of the $x$ axis { (hence representing a situation relevant for the H blow-up case)}, and the last one representing the DA averaged over $10$ angles, only, in the vicinity of the $y$ axis { (hence representing a situation relevant for the V blow-up case)}. Each set of curves includes data from all sixty realisations of the LHC lattices. 

The Beam~1 case features a much smaller spread among the sixty realisations than the Beam~2 case. Apart from this, however, the results for the two beams share a number of common features: the curves representing the average close to the $x$ axis have the mildest dependence on time; the curves representing the average close to the $y$ axis and the global average have a very similar shape featuring almost a constant shift between them. 

{ The numerical model can also provide quantitative information on other lattice properties, such as the detuning with amplitude.} An important condition for the applicability of the approach based on the diffusion equation is that $\partial\Omega/\partial I$ is not too small for the cases under consideration. Indeed, the value of $\partial\Omega/\partial I$ has been evaluated and turned out to be $O(1)$ as requested for the validity of Eq.~\eqref{anglescale}. 

In summary, the results of the symplectic-tracking simulations confirm that the assumptions needed to apply a diffusive approach to the description of beam losses in the LHC are fulfilled. 
\section{Conclusions} \label{sec:conclusions}
In this paper, a novel diffusive model capable of an excellent agreement with the experimental results of the recent dynamic aperture at the CERN LHC has been presented. The model is inspired by the optimal estimate of the remainder of the perturbative series provided by Nekhoroshev theorem, which we propose as functional form of the diffusion coefficient. The model features three parameters that characterise the diffusion equation and the physical meaning of these parameters has been highlighted and discussed in detail. The model has been successfully applied to the description of the beam-loss data sets, which originated from measurements performed at the CERN LHC at $6.5$~TeV. The various data sets represent a number of different configurations in terms of non-linear effects in the beam dynamics, which suggests that the excellent agreement obtained between measurements and the model is a generic feature. One of the model parameters, $\kappa$, is kept the same for all data sets, which is perfectly in line with its physical interpretation. The deviation of the obtained value of $\kappa$ from the theoretical estimate, which is related to the phase space dimension, is being further investigated. As a matter of fact, the theoretical estimates consider a local transport in the action, whereas we are applying the proposed model to a global action diffusion. Therefore, one could expect that in our case $\kappa$ is the result of an effective description of the dynamics on a large spatial scale. Furthermore, it is worth stressing that the results of the numerical simulations depend critically on the values of the model's parameters, i.e. variations at the level of few percent of these parameters change strongly the functional form of the simulated beam losses. This is a very reassuring observation, indicating that the determination of the model parameters is very robust.

In all the considered cases, a 1D approach has been applied and this assumption has been probed by means of symplectic-tracking simulations, which fully supported the choice made. The interesting question whether the diffusion model can be justified on the basis of tracking simulations requires a detailed study of the phase space structure to detect the existence of weakly chaotic regions and to understand the effect of external random perturbations unavoidable in real accelerators. We plan also to investigate an approach based on the 2D Fokker-Planck equation and to examine in depth the relation between the diffusion coefficient and the stability times for the orbits in phase space computed by means of the Nekhoroshev's estimate~\cite{dynap1}. 

It is also worth stressing that the proposed approach can be used to predict the beam losses as well as the evolution of the beam distribution, extrapolating experimental measurements to long time scales that are well beyond the possibility of present symplectic-tracking codes. The optimal approach should be to couple symplectic tracking with the Fokker-Planck equation: the first, performed over a limited number of turns, but with a detailed exploration of phase space, should provide the information about the functional form of the diffusion coefficient; the latter would provide information on the long-term dynamics, such as beam losses and transverse distributions, possibly covering realistic times scales for Accelerator Physics applications. { Therefore, in this way one could apply the outlined approach as a real predictive technique for future high-energy colliders.}
\section*{Acknowledgements}
We would like to express our warm gratitude to the OP and BI groups of the CERN Beams department for the support lent to the experimental studies. One of the authors (AB) would like to thank the CERN Beams department for the hospitality during the work. 
\appendix
\section{Constructing the diffusion equation}\label{sec:app1}
{ The existence of a small positive Ljapounov exponent allows to approximate the orbit's diffusion by means of a stochastic process~\cite{mestre}. Under these conditions, let us consider a generic $2d$-dimensional perturbed Hamiltonian
\begin{equation}
H(\theta,I)=H_0(I)+H_1(\theta,I) \qquad (\theta,I) \in \mathbb{T}^d\times\mathbb{R}^d
\label{pertham}
\end{equation}
where $\Vert H_1(\theta,I)\Vert$ fulfils Nekhoroshev's estimate in a region of phase space where the dynamics is weakly chaotic. The corresponding phase flow satisfies 
\begin{equation}
\begin{array}{lcl}
\theta(T) & = & \theta(0)+\displaystyle{\int_0^T \Omega(I)dt+ \int_0^T \frac{\partial H_1}{\partial I}dt} \\
 & & \\
I(T) & = & I(0)-\displaystyle{\int_0^T \frac{\partial H_1}{\partial \theta}dt}
\label{phasef}
\end{array}
\end{equation}
where $\Omega=\partial H_0/\partial I$ is the phase advance and $T$ represents a time scale sufficiently long with respect to the decorrelation time scale of chaotic orbits, i.e. $T$ is related to the reciprocal of the smallest Ljapounov exponent. We introduce a small parameter $\varepsilon=\Vert H_1\Vert$ as the $L^2$-norm of $H_1$ and we define a new perturbation obtained by normalising the original one, namely $\hat H_1=H_1/\Vert H_1\Vert$. In the limit $\varepsilon\ll 1$ the action dynamics is approximated by a diffusion process~\cite{mestre} at the action-diffusion time scale 
\begin{equation}
T_I\simeq \varepsilon^{-2}\left \Vert \frac{\partial \hat H_1}{\partial \theta} \right \Vert^{-2}
\label{difftime}
\end{equation}
whereas the angle variables tend to relax in a time scale of the order of~\cite{bazzani97}
\begin{equation}
T_\theta \simeq \varepsilon^{-3/2}\left \Vert \frac{\partial \hat H_1}{\partial \theta}\right \Vert^{-3/2} \left \Vert \frac{\partial\Omega}{\partial I}\right \Vert^{-1} \, .
\label{anglescale}
\end{equation}
In the analytic case, if $\varepsilon \ll 1$ and $\partial\Omega/\partial I$ is bounded from below by a quantity $O(1)$, we have $T_\theta \ll T_I$ and the random-phase approximation is justified. 

Then, the diffusion coefficient for the action dynamics is defined as
\begin{equation}
\mathcal{D}_{ij}(I)=\frac{1}{(2\pi)^d}\int_{\mathbb{T}^d} \left (\frac{\partial \hat H_1}{\partial \theta_i} \frac{\partial \hat H_1}{\partial \theta_j}\right )d\theta 
\qquad 1 \leq i,j \leq d \, .
\label{diffcoef}
\end{equation}
and the actions dynamics is approximated by the stochastic differential equation
\begin{equation}
dI_j(\tau) =-\sqrt{\left \langle \frac{\partial \hat H_1}{\partial \theta_j}\frac{\partial \hat H_1}{\partial \theta_l}\right \rangle} \, dw_l(\tau) + \frac{1}{2} \frac{\partial}{\partial I_k}\left \langle \frac{\partial \hat H_1}{\partial \theta_j}\frac{\partial \hat H_1}{\partial \theta_k}\right \rangle d\tau \qquad 1 \leq j, k, l \leq d 
\label{stocaction}
\end{equation}
where summation over repeated indexes has been assumed and $dw_l(\tau), \,\, 1 \leq l \leq n$ are $n$ independent increments of Wiener processes and $\tau=\Vert H_1\Vert^2 t=\varepsilon^2 t$ is the action-diffusion slow time. The physical meaning of the stochastic dynamics~(\ref{stocaction}) is that the weakly-chaotic dynamics can be approximated by a stochastic process at the action-diffusion time scale, where the different realisations depend on the initial conditions, and the action diffusion is so slow that one has a continuous relaxation of the angles variables (see Eq.~\eqref{anglescale}) justifying an averaging principle.

The evolution of a distribution function $\rho_0(I)$ is described by a Fokker-Planck (FP) equation of the form
\begin{equation}
\frac{\partial \rho}{\partial \tau}=
\frac{1}{2}\frac{\partial}{\partial I_i}\mathcal{D}_{ij}(I)
\frac{\partial}{\partial I_j}\rho(I,\tau)
\label{fokker1}
\end{equation}
where $\rho(I,0)=\rho_0(I)$ and boundary conditions of Eq.~\eqref{fokker1} are introduced as absorbing boundaries at $I=I_{\rm abs}$. In this framework, the boundary conditions may represent both physical barriers at a given amplitude, beyond which particles are lost, as well as the frontier of the weakly-chaotic layer where the diffusive approach is fully justified, beyond which fast escape to infinity occurs. It is worthwhile noting that a natural boundary condition at $I=0$ exists, since in the considered model the following limit holds
\begin{equation}
\lim_{I\to 0} \mathcal{D}_{ij}(I)=0 \, .
\end{equation}

In principle, to describe the diffusion behaviour of non-linear betatronic
motion in a circular particle accelerator one needs two action variables $(I_x,I_y)$ that define the non-linear invariants. However, if the diffusion process takes place mainly along a one-dimensional direction, then an approach based on a one degree-of-freedom FP equation is well justified and Eq.~\eqref{fokker1} reduces to the simpler form}
\begin{equation}
\frac{\partial \rho}{\partial \tau}=
\frac{1}{2}\frac{\partial}{\partial I}\mathcal{D}(I)
\frac{\partial}{\partial I}\rho(I,\tau)
\label{FPnew}
\end{equation}
where
$$
\mathcal{D}(I)=\frac{1}{\Vert H_1\Vert^2}\left \langle \left (\frac{\partial H_1}{\partial \theta}\right )^2\right \rangle_\theta
$$
with an absorbing barrier at $I=I_{\rm abs}$.

According to the Nekhoroshev-like estimate for the perturbation term in the Hamiltonian (cfr. Eq.~\eqref{diffnek} in the paper), we assume
\begin{equation}
\mathcal{D}^{1/2}(I)=\sqrt{c}\exp\left [-\left (\frac{I_\ast}{I}\right )^{1/2\kappa}\right ]
\label{diffcoef1}
\end{equation}
where $I_\ast$ depends on the non-linear terms and the exponent $\kappa$ is  related to the dimensionality of the system and $c$ is a normalising constant according to Eq.~(\ref{normal}). In the sequel we set $\alpha=1/(2\kappa)$ to simplify the notation.

We are interested in the evolution of an initial distribution when $I_{\rm abs}\ll I_\ast$. The main problem is to estimate the probability current at the location of the boundary condition and its dependence on the diffusion coefficient. The FP equation~(\ref{FPnew}) can be associated to a stochastic differential equation of the form 
\begin{equation}
dI 
=c \, \frac{\alpha}{I}\left (\frac{I_\ast}{I}\right )^{\alpha}
\exp\left [-2\left (\frac{I_\ast}{I}\right )^{\alpha}\right ]d\tau 
+ \sqrt{c}\, \exp\left [-\left (\frac{I_\ast}{I}\right )^{\alpha}\right ]dw_{\tau} \, .
\label{stoch}
\end{equation}

Introducing the adimensional action $u=I/I_\ast$ we have
\begin{equation}
du 
=c \, \frac{\alpha}{I_\ast^2 \, u^{\alpha+1}}
\exp\left [-2\left (\frac{1}{u}\right )^\alpha\right ]d\tau
+\sqrt{c} \, \frac{1}{I_\ast} \, \exp\left [-\left (\frac{1}{u}\right )^\alpha\right ]dw_{\tau} \, .
\end{equation}
The change of variable
\begin{equation}
x=-\int_u^{u_{\rm abs}}\exp\left [\left (\frac{1}{u}\right )^\alpha\right ]du \qquad u_{\rm abs}=\frac{I_{\rm abs}}{I_\ast}
\label{change}
\end{equation}
reduces the stochastic differential equation~\eqref{stoch} to the form
\begin{equation}
dx=c \, \frac{\alpha}{I_\ast^2} \, a(x) \, d\tau+\sqrt{c} \, \frac{1}{I_\ast}, dw_\tau
\end{equation}
with an absorbing boundary condition at $x=0$ and the drift coefficient given by 
\begin{equation}
a(x)=\frac{1}{u^{\alpha+1}(x)}
\exp\left \{ -\left [\frac{1}{u(x)}\right ]^\alpha\right \}=-\frac{dV}{dx} \, ,
\label{driftn}
\end{equation}
which represents an external forcing towards the boundary that vanishes when $u\to 0$ (i.e. $x\to -\infty$) and it can be associated to the potential $V(x)$.

In the next section we estimate the probability current lost at the absorbing barrier by studying the spectral properties of the Smoluchowski equation
\begin{equation}
\frac{\partial \rho}{\partial t}=\frac{\partial }{\partial x}\frac{dV}{dx}\rho+\mathcal{D} \, \frac{\partial^2\rho}{\partial x^2}
\label{smo}
\end{equation}
when the drift coefficient can be approximated by a constant force and the diffusion coefficient $\mathcal{D}$  is constant. 
\section{Estimate of the probability current at an absorbing boundary for the Smoluchowski equation}\label{sec:app2}
Let us consider the Smoluchowski equation
$$
\frac{\partial \rho}{\partial t}=\frac{\partial }{\partial x}\frac{dV}{dx}\rho+\mathcal{D}\frac{\partial^2\rho}{\partial x^2}
$$
it is possible to cast the diffusion operator in a self-adjoint form by introducing the distribution $p(x,t)$ such that
$$
\rho(x,t)=\exp\left [-\frac{V(x)}{2\, \mathcal{D}}\right ] \, p(x,t) \, ,
$$
which satisfies
\begin{equation}
\frac{\partial p}{\partial t}=\frac{1}{2}\left [\frac{d^2V}{d x^2}-\frac{1}{2\, \mathcal{D}}
\left (\frac{dV}{dx}\right )^2\right ]p+\mathcal{D}\frac{\partial^2 p}{\partial x^2} \, .
\label{fpa}
\end{equation}

The solution of Eq.~\eqref{fpa} can be expanded in the form
$$
p(x,t)=\sum_\lambda c_\lambda(t)\, \phi_\lambda(x)
$$
where
$$
c_\lambda(t)=c_\lambda(0)e^{-\lambda t}
$$
and $\phi_\lambda(x)$ are the eigenfunctions of Eq.~(\ref{fpa}) that satisfy
$$
\frac{1}{\mathcal{D}}\left [a(x)-\lambda\right ]\phi_\lambda(x)=\frac{d^2 \phi_\lambda}{d x^2}
$$
where
$$
a(x)=\frac{1}{2}\left [\frac{1}{2\, \mathcal{D}}\left (\frac{dV}{dx}\right )^2-\frac{d^2V}{d x^2}\right ] \, .
$$

Thanks to the orthogonality and completeness properties, which are expressed by 
\begin{align}
\int \phi_\lambda(x)\phi_{\lambda'}(x) dx & =\delta(\lambda-\lambda')  \notag \\
& \\
\sum_{\lambda} \phi_\lambda(x)\phi_\lambda(x') & =\delta(x-x') \, , \notag
\end{align}
and using the initial conditions
\begin{align}
\rho(x,0)& =\exp\left [-\frac{V(x)}{2\, \mathcal{D}}\right ] p(x,0) \notag \\
& \\
p(x,0) & = \sum_\lambda c_\lambda(0)\phi_\lambda(x) \notag 
\end{align}
one obtains that 
$$
c_{\lambda}(0)=\int \exp\left [\frac{V(x)}{2\, \mathcal{D}}\right ] \rho(x,0)\phi_{\lambda}(x) dx \, .
$$
Whenever $\rho(x,0)=\delta(x-x_0)$ the solution reads
$$
\rho(x,t)=\exp\left [-\frac{V(x)-V(x_0)}{2\, \mathcal{D}}\right ]\sum_\lambda e^{-\lambda t}\phi_\lambda(x_0)\phi_\lambda(x) \, .
$$

If we set an absorbing boundary condition at $x=0$, i.e. $\rho(0,t)=0$, then we require $\phi_\lambda(0)=0$  and the probability current at the boundary reads
\begin{equation}
J(t) 
=\mathcal{D} \left . \frac{\partial \rho}{\partial x} \right \vert_{(0,t)} =\mathcal{D} \, \exp\left [-\frac{(V(0)-V(x_0)}{2\, \mathcal{D}}\right ] 
\times \sum_\lambda e^{-\lambda t}\phi_\lambda(x_0)
\left . \frac{d \phi_\lambda}{d x} \right \vert_{(0)} \, .
\label{currsh}
\end{equation}

Let us consider the special case $V(x)=-\nu x$, i.e. a constant drift towards the boundary with an absorbing  condition at $x=0$, then $a(x)=\nu^2/(4\, \mathcal{D}) $ and the self-adjoint operator  
$$
-\frac{\nu^2}{4\, \mathcal{D}}+\mathcal{D}\frac{\partial^2 }{\partial x^2}
$$
is negative defined, so that all eigenvalues satisfy $\lambda \le 0$. Indeed, $\mathcal{D}\, \partial^2/\partial x^2$ is also a negative defined symmetric operator and its spectrum (except for the zero eigenvalue) is bounded from above by $-\nu^2/(4\, \mathcal{D})$. Let $-\lambda$ be the eigenvalue, the eigenvector equation reads 
$$
-\frac{1}{\mathcal{D}}\left [\lambda-\frac{\nu^2}{4\, \mathcal{D}}\right ]\phi_\lambda(x)=\frac{d^2 \phi_\lambda}{d x^2}
$$
with the boundary condition $\phi_\lambda(0)=0$. If we set
$$
k_\lambda=\sqrt{\lambda -\frac{\nu^2}{4\, \mathcal{D}}}
$$
we have non-trivial solutions given by 
$$
\phi_\lambda(x)=\frac{1}{\sqrt{\pi}}\sin\left(\frac{k_\lambda}{\sqrt{\mathcal{D}}} x\right ) \, .
$$

The zero-eigenvalue corresponds to a trivial solution and we have an upper limit for the negative eigenvalues from the condition
$$
\lambda -\frac{\nu^2}{4\, \mathcal{D}}> 0\qquad \Rightarrow \qquad \lambda_{\rm min}=\frac{\nu^2}{4\, \mathcal{D}}
$$
$\lambda_{\rm min}^{-1}$ defines the decaying characteristic time for the Fiedler's eigenvector. Therefore, the existence of a constant drift field implies an upper bound to the Fiedler's eigenvalue. 

As an example, we compute the probability current at the absorbing barrier $x=0$ for the case $\nu=0$ so that $\phi_\lambda(x)\propto\sin\sqrt{\lambda/\mathcal{D}} \, x$. From Eq.~(\ref{currsh}) with $\delta(x-x_0)$ as initial condition and for a continuous spectrum, we have 
\begin{equation}
J(t, x_0) 
=\frac{\mathcal{D}}{\pi}\int_0^\infty e^{-\lambda t}\sin\left(\sqrt{\frac{\lambda}{\mathcal{D}}}x_0\right )\sqrt{\frac{\lambda}{\mathcal{D}}}d\left (\sqrt{\frac{\lambda}{\mathcal{D}}}\right )
= \frac{\mathcal{D}}{\pi}\int_0^\infty e^{-\mu^2 \, \mathcal{D} \, t}\sin\left(\mu \, x_0\right ) \, \mu \, d \mu
\end{equation}
where $\lambda=\mu^2\, \mathcal{D}$. The integral reduces to a Gaussian integral since
\begin{equation}
\exp\left (-\mu^2 \, \mathcal{D} \, t \pm i \, \mu \, x_0\right) 
=\exp\left [-\mathcal{D}\, t\, \left (\mu \pm i\frac{x_0}{2\, \mathcal{D}\, t}\right )^2\right ] 
\times \exp\left (-\frac{x_0^2}{4\, \mathcal{D}\, t}\right )
\end{equation}
and
\begin{equation}
\begin{split}
\int_0^\infty \left \{ \exp\left [-\mathcal{D}\, t\left (\mu - i\frac{x_0}{2\, \mathcal{D}\, t}\right )^2\right ] - \exp\left [-\mathcal{D}\, t\left (\mu + i\frac{x_0}{2\, \mathcal{D}\, t}\right )^2\right ] \right \} \mu \, d \mu  & =\int_{-\infty}^\infty \exp\left [-\mathcal{D}\, t\left (\mu- i\frac{x_0}{2\, \mathcal{D}\, t}\right )^2\right ] \mu \, d \mu = \\ 
& = i \, \frac{x_0}{2}\sqrt{\frac{\pi}{(\mathcal{D}\, t)^{3}}}  \, . 
\end{split}
\end{equation}
Then we obtain
\begin{equation}
J(t, x_0) 
= \frac{\mathcal{D}}{2i \, \pi}\exp\left (-\frac{x_0^2}{4\, \mathcal{D}\, t}\right ) 
 \int_{-\infty}^\infty \exp\left [- \mathcal{D}\, t \left (\mu- i\frac{x_0}{2\, \mathcal{D}\, t}\right )^2\right ] \mu \, d \mu 
= \frac{x_0}{\sqrt{2\, \pi \, \mathcal{D}}(2\, t)^{3/2}} \exp\left (-\frac{x_0^2}{4\, \mathcal{D}\, t}\right ) \notag \, . 
\end{equation}

The presence of a drift field modifies the previous calculation and if we define $\lambda-\nu^2/(4\, \mathcal{D})=k_\lambda^2=\mu^2\, \mathcal{D}$ we have the relation
\begin{equation}
\lambda \, t\pm i\frac{k_\lambda}{\sqrt{\mathcal{D}}}x_0=\left (\mu^2\, \mathcal{D}+\frac{\nu^2}{4\, \mathcal{D}}\right )t\pm i \, \mu \, x_0= 
\frac{\nu^2}{4\, \mathcal{D}}\, t 
+ \mathcal{D}\, t\left (\mu \pm i\frac{x_0}{2\, \mathcal{D}\, t}\right )^2+\frac{x_0^2}{4\, \mathcal{D}\, t}
\end{equation}
from which we get the Gaussian
$$
\exp\left [-\frac{\nu^2 \, t}{4\, \mathcal{D}}-\frac{x_0^2}{4\, \mathcal{D}\, t}\right ] \exp\left [-\mathcal{D}\, t\left (\mu \pm i\frac{x_0}{2\, \mathcal{D}\, t}\right )^2\right ] \, .
$$
and the probability current reads
\begin{equation}
J(t, x_0) =\frac{x_0 \, \exp\left (-\displaystyle{\frac{x_0 \, \nu}{2\, \mathcal{D}}}\right )}{\sqrt{16\, \pi \, \mathcal{D} \, t^3}} \, \exp\left [-\frac{(x_0+\nu \, t)^2}{4\, \mathcal{D}\, t}\right ]  \, .
\end{equation}
In this case, i.e. for a $\delta$-function initial distribution and a linear drift field, the probability current is described by a Gaussian function moving towards the absorbing barrier and which is diffusing. In adiabatic regime it is possible to approximate the probability current for a slowly-varying drift field according to 
\begin{equation}
J(t, x_0)\simeq \frac{x_0 \, \exp\left (-\displaystyle{\frac{x_0 \, \nu}{2\, \mathcal{D}}}\right )}{\sqrt{16\, \pi \, \mathcal{D} \, t^3}} \, \exp\left \{-\frac{[\Phi^t(x_0)]^2}{4\, \mathcal{D}\, t}\right \}
\label{currentJ}
\end{equation}
where $\Phi^t(x_0)$ is the phase flow of the drift force, namely
\begin{equation}
\frac{dx}{dt}=-\frac{dV}{dx} \, .
\end{equation}
For a generic initial distribution $\rho_0(x)$ we have
\begin{equation}
J(t) \simeq \frac{1}{\sqrt{16\, \pi \, \mathcal{D} \, t^3}} \int_0^\infty x \, \exp \left (-\displaystyle{\frac{x \, \nu}{2\, \mathcal{D}}} \right ) \exp \left \{-\frac{\left [\Phi^t(x) \right ]^2}{4\, \mathcal{D}\, t}\right \} \rho_0(x) \, dx \, .
\end{equation}
\section{Estimate of the probability current at an absorbing barrier with a Nekhoroshev-like diffusion coefficient}\label{sec:app3}
For a Nekhoroshev-like estimate of the diffusion coefficient, we perform the change of variable (cfr. Eq.~(\ref{change}))
\begin{equation}
x=-\int_u^{u_{\rm abs}} \exp\left (u^{-\alpha} \right ) du\simeq (u-u_{\rm abs})\exp\left (u^{-\alpha} \right )
\end{equation}
and we use the diffusion-scaled time $\tau \to \tau \, c /I_\ast^2$. The drift field for the new FP equation is then given by 
\begin{equation}
\frac{dx}{d\tau}=\alpha \, u^{-(\alpha+1)}\exp\left (-u^{-\alpha} \right )\simeq \alpha \, u_0^{-(\alpha+1)}\exp\left (-u_0^{-\alpha} \right )
\label{drift}
\end{equation}
for an initial distribution centred at $u_0$, where an average field approximation has been assumed. Then from Eq.~(\ref{currentJ}) we get the expression for the out-going probability current in the initial time
\begin{equation}
J(t, x_0) \simeq \frac{2\, x_0 \, I_\ast^3}{\sqrt{\pi} \, c^{3/2}\, \varepsilon^3 \, t^{3/2}} \, 
\exp\left [-\frac{\beta(t)}{4 \, I_\ast^2 \, c \, \varepsilon^2 \, t}\right ]
\label{currentnek}
\end{equation}
where 
\begin{equation}
\sqrt{\beta(t)} 
= 2\, x_0 \, I_\ast^2+\alpha \left (\frac{I_0}{I_\ast}\right)^{-(\alpha+1)} 
\times \exp\left [-\left (\frac{I_0}{I_\ast} \right )^{-\alpha} \right ] \, c \, \varepsilon^2 \, t \, .
\end{equation}
$I_0$ and $x_0(I_0)$ define the initial distribution $\delta(I-I_0)$ in the action variable and the $x$ variable, respectively (cfr. Eq.~(\ref{change})). The expression~(\ref{currentnek}) underestimates the probability current at the absorbing boundary since the drift term~(\ref{driftn}) is  a positive-increasing function if the initial condition satisfies $I_0\ll I_\ast$. Indeed, a crude estimate gives
$$
x_0^2\simeq \left( \frac{I_{\rm abs}}{I_\ast}\right )^2\exp\left [2\left (\frac{I_\ast}{I_0}\right )^{\alpha} \right ]
$$
in the case of a Nekhoroshev-like diffusion coefficient. Moreover, we expect that the probability current decreases exponentially in the initial condition $I_0$
$$
J \propto \left( \frac{I_\ast}{I_{\rm abs}}\right )^2\exp\left [2\left (\frac{I_\ast}{I_0}\right )^{\alpha} \right ] \, .
$$

The time to lose a fixed probability fraction $\Delta P$ starting from an initial condition $\delta(I-I_0)$ is computed by integrating the probability current, namely
\begin{equation}
\Delta P 
=\int_0^{t_{\rm DA}} J(t,x_0)dt\simeq x_0^{-2} 
\times \int_0^{x_{\Delta P}} \frac{x^3}{4\sqrt{\pi}}
\exp\left(-\frac{x^2}{2}\right )\frac{dt}{dx} dx 
=-\frac{I_\ast^2}{2\, \sqrt{\pi} \, c \, \varepsilon^2}\int_0^{x_{\Delta P}} x\, \exp\left(-\frac{x^2}{2}\right )dx
\end{equation}
where we neglect the drift-term contribution, which is justified by the exponential decaying of the drift field (\ref{drift}), and we use the auxiliary variable $x$ defined as
\begin{equation}
t=\frac{x_0^2 \, I_\ast^2}{c\, \varepsilon^2 \, x} \qquad \Rightarrow \qquad \frac{dt}{dx}=-\frac{x_0^2 \, I_\ast^2}{c \, \varepsilon^2 \, x^2} \, .
\label{deft}
\end{equation}
Then we obtain a relationship between $\Delta P$ and $x_{\Delta P}$, i.e.
\begin{equation}
\Delta P=-\frac{I_\ast^2}{2\, \sqrt{\pi} \, c \, \varepsilon^2}\left [1-\exp\left(-\frac{x_{\Delta P}^2}{2}\right )\right ]
\label{currentotn}
\end{equation}
and we estimate the dynamic aperture timescale $t_{\rm DA}$ from the definition (\ref{deft}), letting $x_{\Delta P} \simeq 1$ in Eq.~(\ref{currentotn})
\begin{equation}
t_{\rm DA}=\frac{x_0^2 \, I_\ast^2}{c \, \varepsilon^2}\simeq \frac{I_{\rm abs}^2}{c \, \varepsilon^2} \,
\exp\left [2\left (\frac{I_\ast}{I_0}\right )^{\alpha} \right ] \, . 
\label{dynap}
\end{equation}

Therefore $t_{\rm DA}$ turns out to be exponentially long as we change the initial condition $I_0$ in agreement with the tracking results~\cite{dynap1,dynap2}. 
\end{document}